\magnification=1200
\null\bigskip\bigskip
\def\qed{\unskip\kern 6pt\penalty 500\raise -2pt\hbox
{\vrule\vbox to 10pt{\hrule width 4pt\vfill\hrule}\vrule}}
\centerline{SMOOTH DYNAMICS AND}
\centerline{NEW THEORETICAL IDEAS IN NONEQUILIBRIUM STATISTICAL MECHANICS.}
\bigskip
\centerline{by David Ruelle\footnote{*}{IHES., Bures sur Yvette, 
$<$ruelle@ihes.fr$>$, and Math. Dept., Rutgers University.}}
\bigskip\bigskip\noindent
	{\sl Abstract.  This paper reviews various applications of the theory of smooth dynamical systems to conceptual problems of nonequilibrium statistical mecanics.  We adopt a new point of view which has emerged progressively in recent years, and which takes seriously into account the chaotic character of the microscopic time evolution.  The emphasis is on {\it nonequilibrium steady states} rather than the traditional {\it approach to equilibrium} point of view of Boltzmann.  The nonequilibrium steady states, in presence of a Gaussian thermostat, are described by {\it SRB measures}.  In terms of these one can prove the Gallavotti-Cohen {\it fluctuation theorem}.  One can also prove a general {\it linear response formula} and study its consequences, which are not restricted to near equilibrium situations.  At equilibrium one recovers in particular the {\it Onsager reciprocity relations}.  Under suitable conditions the nonequilibrium steady states satisfy the {\it pairing theorem} of Dettmann and Morriss.  The results just mentioned hold so far only for classical systems; they do not involve large size, {\it i.e.} they hold without a thermodynamic limit.}
\vfill\eject
\centerline{Chapter 1. A GENERAL DISCUSSION.}
\bigskip\bigskip
	{\bf 1. Introduction.}
\medskip
	Statistical mechanics, which was created at the end of the 19-th century by such people as Maxwell, Boltzmann, and Gibbs, consists of two rather different parts: equilibrium and nonequilibrium statistical mechanics.  The success of equilibrium statistical mechanics has been spectacular.  It has been developed to a high degree of mathematical sophistication, and applied with success to subtle physical problems like the study of critical phenomena.  Equilibrium statistical mechanics also has highly nontrivial connections with the mathematical theory of smooth dynamical systems and the physical theory of quantum fields.
\medskip
	By contrast, the progress of nonequilibrium statistical mechanics has been much slower.  We still depend on the insights of Boltzmann for our basic understanding of irreversibility, and this understanding remains rather qualitative.  Further progress has been mostly on dissipative phenomena close to equilibrium: Onsager reciprocity relations, Green-Kubo formula, and related results.  Yet, there is currently a strong revival of nonequilibrium statistical mechanics, based on taking seriously the complications of the underlying microscopic dynamics.  The purpose of the present review paper is to discuss some of the new ideas in the dynamical approach to the problems of nonequilibrium.  For a different presentation, see [28].
\medskip
	This paper has grown out of lectures given in the Mathematics Department at Rutgers University in 1997 and 1998, where I benefitted from a very interactive and exceptionally competent audience.  I have thus used in what follows what I learnt from discussions with Joel Lebowitz, Giovanni Gallavotti, Sheldon Goldstein, Eugene Speer, Michael Kiesling, Frederico Bonetto, and others.  Let all of them be thanked.
\medskip
	{\bf 2. Irreversibility.}
\medskip
	The rest of this chapter is devoted to a discussion of irreversibility based on Boltzmann's ideas.  This will serve as an introduction and background for our later discussions.
\medskip
	It is a fact of common observation that the behavior of bulk matter is often irreversible: if a warm and a cold body are put in contact, they will equalize their temperatures, but once they are at the same temperature they will not spontaneously go back to the warm and cold situation.  Such facts have been organized into a body of knowledge named {\it thermodynamics}.  According to thermodynamics, a number called {\it entropy} can be associated with macroscopic systems which are (roughly speaking) locally in equilibrium.  The definition is such that, when the system is isolated, its entropy can only increase with time or stay constant.  (This is the celebrated {\it second law of thermodynamics}).  A strict increases in entropy corresponds to an irreversible process.  Such processes are also called {\it dissipative}, because they dissipate noble forms of energy (like mechanical energy) into heat.  The flow of a viscous fluid, the passage of an electric current through a conductor, or a chemical reaction like a combustion are typical dissipative phenomena.  The purpose of nonequilibrium statistical mechanics is to explain irreversibility on the basis of microscopic dynamics, and to give quantitative predictions for dissipative phenomena.  In what follows we shall assume that the microscopic dynamics is classical.  Unfortunately we shall have little to say about the quantum case.
\medskip
	{\bf 3. Dynamics and entropy.}
\medskip
	We shall begin with a naive discussion, and then see how we have to modify it to avoid the difficulties that will arise.  The microscopic time evolution $(f^t)$ which we want to consider is determined by an evolution equation
$$	{dx\over dt}=F(x)\eqno{(1)}     $$
in a phase space ${\cal M}$.  More precisely, for an isolated system with Hamiltonian $H$, we may rewrite (1) as
$$	{d\over dt}\pmatrix{{\bf p}\cr{\bf q}\cr}
	=\pmatrix{-\partial_{\bf q}H({\bf p},{\bf q})\cr
	\partial_{\bf p}H({\bf p},{\bf q})\cr}\eqno{(2)}     $$
where ${\bf p}$, and ${\bf q}$ are $N$-dimensional for a system with $N$ degrees of freedom (${\cal M}$ is thus $2N$-dimensional).  Note that we are interested in a macroscopic description of a macroscopic system where $N$ is large, and microscopic details cannot be observed.  In fact many different points $({\bf p},{\bf q})$ in phase space have the same macroscopic appearance.  It appears therefore reasonable to describe a macroscopic state of our system by a probability measure $m(dx)$ on ${\cal M}$.  At equilibrium, $m(dx)$ should be an $(f^t)$-invariant measure.
\medskip
	Remember now that the Hamiltonian time evolution determined by (2) preserves the energy and the volume element $d{\bf p}\,d{\bf q}$ in phase space.  This leads to the choice of the {\it microcanonical ensemble}
$$	m_K(d{\bf p}\,d{\bf q})
={1\over\Omega_K}\delta(H({\bf p},{\bf q})-K)d{\bf p}\,d{\bf q}\eqno{(3)}  $$
to describe the equilibrium state of energy $K$.  [$\Omega_K$ is a normalization constant, and the {\it ergodic hypothesis} asserts that $m_K$ is ergodic].  Note that the support of $m_K$ is the energy shell
$$	{\cal M}_K=\{({\bf p},{\bf q}):H({\bf p},{\bf q})=K\}     $$
\par
	If the probability measure $m(d{\bf p}\,d{\bf q})$ has density $\underline m({\bf p},{\bf q})$ with respect to $d{\bf p}\,d{\bf q}$, we associate with the state described by this measure the entropy
$$S(\underline m)=-\int d{\bf p}\,d{\bf q}\,\underline m({\bf p},{\bf q})
	\log\underline m({\bf p},{\bf q})\eqno{(4)}     $$
This is what Lebowitz [62] calls the {\it Gibbs entropy}, it is the accepted expression for the entropy of equilibrium states\footnote{*}{There is a minor technical complication here.  While (4) gives the right result for the {\it canonical ensemble} one should, for the {\it microcanonical ensemble}, replace the reference measure $d{\bf p}\,d{\bf q}$ in (3) by $\delta(H({\bf p},{\bf q})-K)d{\bf p}\,d{\bf q}$.  This point is best appreciated in the light of the theory of equivalence of ensembles in equilibrium statistical mechanics.  For our purposes, the easiest is to replace $\delta$ in (3) by the characteristic function of $[0,\epsilon]$ for small $\epsilon$.  The measure $m_K$ is then still invariant, but no longer ergodic.  Note also that the traditional definition of the entropy has a factor $k$ (Bolzmann's constant) in the right-hand side of (4).}, and it remains reasonable outside of equilibrium, as information theory for instance would indicate.  Writing $x$ instead of $({\bf p},{\bf q})$, we know that the density of $\widehat{m}=f^{t*}m$ is given by
$$  \widehat{\underline{m}}(x)={\underline{m}(f^{-t}x)\over J_t(f^{-t}x)}  $$
where $J_t$ is the Jacobian determinant of $f^t$.  We have thus
$$ S(\underline{m})=-\int dx\,\underline{m}(x)\log\underline{m}(x)\eqno{(5)} $$
$$	S(\widehat{\underline{m}})=-\int dx\,\widehat{\underline{m}}(x)
\log\widehat{\underline{m}}(x)=-\int dx\,{\underline{m}(f^{-t}x)\over
	J_t(f^{-t}x)}\log{\underline m(f^{-t}x)\over J_t(f^{-t}x)}     $$
$$	=-\int dx\,\underline{m}(x)\log{\underline{m}(x)\over J_t(x)}
	=S(\underline{m})+\int dx\,\underline{m}(x)\log J_t(x)\eqno{(6)}     $$
In the Hamiltonian situation that we are considering for the moment, since the volume element is preserved by $f^t$, $J_t=1$, and therefore $S(\widehat{\underline{m}})=S(\underline{m})$.
\medskip
	So, we have a problem: the entropy seems to remain constant in time.  In fact, we could have expected that there would be a problem because the Hamiltonian evolution (2) has time reversal invariance (to be discussed later), while we want to prove an increase of entropy which does not respect time reversal.  We shall now present Boltzmann's way out of this difficulty.
\medskip
	{\bf 4. The classical ideas of Boltzmann and his followers.}
\medskip
	Let us cut ${\cal M}$ into cells $c_i$ so that the coordinates ${\bf p}$, ${\bf q}$ have roughly fixed values in a cell.  In particular all points in $c_i$ are macroscopically equivalent (but different $c_i$ may be macroscopically indistinguishable).  Instead of describing a macroscopic state by a probability measure $m$, we may thus give the weights $m(c_i)$.  Now, time evolution will usually distort the cells, so that each $f^tc_i$ will now intersect a large number of cells. If the initial state $m$ occupies ${\cal N}$ cells with weights $1/{\cal N}$ (taken to be equal for simplicity), the state $f^{*t}m$ will occupy (thinly) ${\cal N}^t$ cells with weights $1/{\cal N}^t$, where ${\cal N}^t$ may be much larger than ${\cal N}$.  If the $c_i$ have side $h$, we have
$$	\log{\cal N}=-\sum_im(c_i)\log m(c_i)
	\approx-\int dx\,{\underline{m}}(x)\log({\underline{m}}(x)
	\ldotp h^{2N})=S(\underline m)-2N\log h     $$
{\it i.e.}, the entropy associated with $m$ is roughly $\log{\cal N}+2N\log h$.  The {\it apparent} entropy associated with $f^{t*}m$ is similarly $\log{\cal N}^t+2N\log h$; it differs from $S(\widehat{\underline{m}})$ because the density $\widehat{\underline{m}}$, which may fluctuate rapidly in a cell, has been replaced by an average for the computation of the {\it apparent} entropy.  The entropy increase $\log{\cal N}^t-\log{\cal N}$ is due to the fact that the initial state $m$ is concentrated in a small region of phase space, and becomes by time evolution spread (thinly) over a much larger region.  The time evolved state, after a little smoothing ({\it coarse graining}), has a strictly larger entropy than the initial state.  This gives a microscopic interpretation of the second law of thermodynamics.  In specific physical examples (like that of two bodies in contact with initially different temperatures) one sees that the time evolved state has correlations (say between the microscopic states of the two bodies, after their temperatures have equalized) which are macroscopically unobservable.  In the case of a macroscopic system locally close to equilibrium (small regions of space have a definite temperature, pressure,\dots) the above classical ideas of Boltzmann have been expressed in particularly clear and compelling manner by J. L. Lebowitz [62].  He defines a local {\it Boltzmann entropy} to be the equilibrium entropy corresponding to the temperature, pressure,\dots which approximately describe locally a nonequilibrium state.  The integral over space of the Boltzmann entropy is then what we have called apparent entropy, and is different from the {\it Gibbs entropy} defined by (4).  These ideas would deserve a fuller discussion, but here we shall be content to refer to [62], and to Boltzmann's original works.
\medskip
	There is still some opposition to Boltzmann's ideas (notably by I. Prigogine and his school), but most workers in the area accept them, and so shall we.  We shall however develop a formalism which is both rather different from and completely compatible with the ideas of Boltzmann and Lebowitz just discussed.
\medskip
	{\bf 5. The description of states by probability measures.}
\medskip
	In the above discussion, we have chosen to describe states by probability measures.  One may object that the state of a (classical) system is represented by a point in phase space rather than by a probability measure.  But for a many-particle system like those of interest in statistical mechanics it is practically impossible to know the position of all the particles, which changes rapidly with time anyway.  The information that we have is macroscopic or statistical\footnote{*}{Information on local microscopic correlations is obtained by diffraction techniques.} and it is convenient to take as description of our system a probability distribution $\rho$ on phase space compatible with the information available to us.  Trying to define this $\rho$ more precisely at this stage leads to the usual kind of difficulties that arise when one wants to obtain from a definition what should really come as a result of theory.
\medskip
	For our later purposes it is technically important to work with states described by probability measures.  Eventually, we shall obtain results about individual points of phase space (true almost everywhere with respect to some measure).  From a physical point of view, it would be desirable to make use here of points of phase space which are {\it typical} for a macroscopic state of our system.  Unfortunately, a serious discussion of this point seems out of reach in the present state of the theory.
\medskip
	{\bf 6. Beyond Boltzmann.}
\medskip
	There are two reasons why one would want to go beyond the ideas presented above.  One concerns explicit calculations, like that of a rate of entropy production; the other concerns situations far from equilibrium.  We discuss these two points successively.
\medskip
	If one follows Boltzmann and uses a decomposition of phase space into cells to compute entropy changes, the result need not be monotone in time, and will in general depend on the particular decomposition used.  Only after taking a limit $t\to\infty$ can one let the size of cells tend to 0, and obtain a result independent of the choice of coarse graining.  This leaves open the problem of computing the entropy production per unit time.
\medskip
	The idea of using local Boltzmann entropies works only for a macroscopic system locally close to equilibrium, and one may wonder what happens far from equilibrium.  In fact one finds statements in the literature that biological processes are far from equilibrium (which is true), and that they may violate the second law of thermodynamics (which is not true).  To see that life processes or other processes far from equilibrium cannot violate the second law, we can imagine a power plant fueled by sea water: it would produce electric current and reject colder water in the sea.  Inside the plant there would be some life form or other physico-chemical system functioning far from equilibrium and violating the second law of thermodynamics.  The evidence is that this is impossible, even though we cannot follow everything that happens inside the plant in terms of Boltzmann entropies of systems close to equilibrium.  In fact, Boltzmann's explanation of irreversibility reproduced above applies also here, and the only unsatisfactory feature is that we do not have an effective definition of entropy far from equilibrium. 
\medskip
	The new formalism which we shall introduce below is quite in agreement with the ideas of Boltzmann which we have described.  We shall however define the physical entropy by (4) (Gibbs entropy).  To avoid the conclusion that this entropy does not change in time for a Hamiltonian time evolution, we shall idealize our physical setup differently, and in particular introduce a thermostat.
\medskip
	{\bf 7. Thermostats.}
\medskip
	Instead of investigating the approach to equilibrium as we have done above following Boltzmann, we can try to produce and study nonequilibrium steady states.  To keep a finite system outside of equilibrium we subject it to non-Hamiltonian forces.  We consider thus an evolution of the form (1), but not (2).  Since we no longer have conservation of energy, $x(t)$ cannot be expected to stay in a bounded region of phase space.  This means that the system will heat up.  Indeed, this is what is observed experimentally: dissipative systems produce heat.  An experimentalist will eliminate excess heat by use of a thermostat, and if we want to study nonequilibrium steady states we have to introduce
the mathematical equivalent of a thermostat.
\medskip
	In the lab, the system in which we are interested (called {\it small system}) would be coupled with a {\it large system} constituting the thermostat.  The obvious role of the large system is to take up the heat produced in the small system.  At the same time, the thermostat allows entropy to be produced in the small system by a mechanism discussed above: microscopic correlations which exist between the small and the large system are rendered unobservable by the time evolution.  An exact study of the pair small+large system would lead to the same problems that we have met above with Boltzmann's approach\footnote{*}{For such an approach, see Jak\v si\'c and Pillet [53], where the large system has infinitely many noninteracting degrees of freedom.  Studying the {\it diffusion and loss of correlations} in an infinite interacting system (say a system of particles with Hamiltonian interactions) appears to be very difficult in general, because the same particles may interact again and again, and it is hard to keep track of the correlations resulting from earlier interactions.  This difficulty was bypassed by Lanford when he studied the Boltzmann equation in the Grad limit [60], because in that limit, two particles that collide once will never see each other again.}.  Note however that the thermostats used in the lab are such that their state changes as little as possible under the influence of the
small system.  For instance the small system will consist of a small amount of fluid, surrounded by a big chunk of copper constituting the large system: because of the high thermal conductivity of copper, and the bulk of the large system, the temperature at the fluid-copper interface will remain constant to a good precision.  In conclusion, the experimentalist tries to build an {\it ideal thermostat}, and we might as well do the same.  We replace thus (1) by
$$	{dx\over dt}=F(x)+\Theta(\omega(t),x)     $$
where the effect $\Theta(\omega(t),x)$ of the thermostat depends on the state $x$ of the small system, and on the state $\omega(t)$ of the thermostat, but the time evolution $t\to\omega(t)$ does not depend on the state $x$ of the small system.
\medskip
	We may think of $\omega(t)$ as random (corresponding to a {\it random thermostat}), but the simplest choice is to take $\omega$ constant, and use $\Theta(x)=\Theta(\omega,x)$ to keep $x(t)$ on a compact manifold $M$.  For instance if $M$ is the manifold $\{x:h(x)=K\}$, we may take
$$	\Theta(x)
=-{(F(x)\,\cdotp{\rm grad}h(x))\over({\rm grad}h(x)\,\cdotp{\rm grad}h(x))}
	\,{\rm grad}h(x)     $$
This is the so-called {\it Gaussian thermostat} [51], [32].  We shall be particularly interested later in the {\it isokinetic thermostat}, which is the special case where $x=({\bf p},{\bf q})$ and $h(x)={\bf p}\cdotp{\bf p}/2m$ (kinetic energy).
\medskip
	{\bf 8. Nonequilibrium steady states.}
\medskip
	We assume for the moment that the phase space of our system is reduced by the action of a thermostat to be a compact manifold $M$.  The time evolution equation on $M$ has the same form as (1):
$$	{dx\over dt}=F(x)\eqno{(7)}     $$
where the vector field $F$ on $M$ now describes both the effect of nonhamiltonian forces and of the thermostat.  Note that (7) describes a general smooth evolution, and one may wonder if anything of physical interest is preserved at this level of generality.  Perhaps surprisingly, the answer is yes, as we see when we ask what are the physical stationary states for (7).
\medskip
	We start with a probability measure $m$ on $M$ such that $m(dx)=\underline{m}(x)\,dx$, where $dx$ is the volume element for some Riemann metric on $M$ (for simplicity, we shall say that $m$ is {\it absolutely continuous}).  At time $t$, $m$ becomes $(f^t)^*m$, which still is absolutely continuous.  If $(f^t)^*m$ has a limit $\rho$ when $t\to\infty$, then $\rho$ is invariant under time evolution, and in general {\it singular} with respect to the Riemann volume element $dx$.  [A time evolution of the form (7) does not in general have an absolutely continuous invariant measure].  The probability measures
$$	\rho=\lim_{t\to\infty}(f^t)^*m     $$
or more generally
$$  \rho=\lim_{t\to\infty}{1\over t}\int_0^t d\tau\,(f^\tau)^*m\eqno{(8)}  $$
(with $m$ absolutely continuous) are natural candidates to describe nonequilibrium stationary states, or {\it nonequilibrium steady states}.  Examples of such measures are the {\it SRB states} (see [92], [79], [14], [66], [67]), which will be discussed in greater detail in Chapter 2.
\medskip
	{\bf 9. Entropy production.}
\medskip
	We return now to our calculation (6) of the entropy production:
$$	S(\widehat{\underline{m}})-S(\underline{m})
	=\int dx\,\underline{m}(x)\log J_t(x)     $$
where $\widehat{\underline{m}}$ is the density of $\widehat{m}=f^{t*}m$.  This is the amount of entropy gained by the system under the action of the external forces and thermostat in time $t$.  The amount of entropy produced by the system and given to the external world in one unit of time is thus (with $J=J_1$)
$$	e_f(m)=-\int m(dx)\log J(x)     $$
Notice that this expression makes sense also when $m$ is a singular measure.  The average entropy production in $t$ units of time is
$$	{1\over t}\sum_{k=0}^{t-1}e_f(f^km)
	={1\over t}[S(\underline{m})-S(\widehat{\underline{m}})]\eqno{(9)}   $$
When $t\to\infty$, this tends according to (8) towards
$$	e_f(\rho)=-\int\rho(dx)\,\log J(x)\eqno{(10)}     $$
which is thus the entropy production in the state $\rho$ (more precisely, the entropy production per unit time in the nonequilibrium steady state $\rho$).  Using (7) we can also write
$$	e_f(\rho)=-\int\rho(dx)\,{\rm div}F(x)\eqno{(11)}     $$
\medskip
	Notice that the entropy $S$ in (9) is bounded above (see the definition (5)), so that $e_f(\rho)\ge0$, and in many cases $e_f(\rho)>0$ as we shall see later.  Notice that for an arbitrary probability measure $\mu$ (invariant or not), $e_f(\mu)$ may be positive or negative, but the definition (8) of $\rho$ makes a choice of the direction of time, and results in positive entropy production.  It may appear paradoxical that the state $\rho$, which does not change in time, constantly gives entropy to the outside world.  The solution of the paradox is that $\rho$ is (normally) a singular measure and therefore has entropy $-\infty$: the nonequilibrium steady state $\rho$ is thus a bottomless source of entropy.
\medskip
	{\bf 10. A new idealization of nonequilibrium processes.}
\medskip
	We have now reached a new framework idealizing nonequilibrium processes.  Instead of following Boltzmann in his study of {\it approach to equilibrium}, we try to understand the {\it nonequilibrium steady states} as given by (8).  For the definition of the entropy production $e_f(\rho)$ we use the rate of phase space contraction (10) or (11).  (An early reference proposing this definition is Andrey [1]).  In Chapter 2, we shall discuss the SRB states, which provide a mathematically precise definition of the {\it nonequilibrium steady states}.  After that, the idea is to use SRB states to make interesting physical predictions, making hyperbolicity assumptions (this will be explained later) as strong as needed to get results.  Such a program was advocated early by Ruelle, but only recently were interesting results actually obtained, the first one being the {\it fluctuation theorem} of Gallavotti and Cohen [44], [45].  This will be discussed in Chapter 3.  In later chapters we shall review the growing body of knowledge which is emerging in this new approach to nonequilibrium statistical mechanics.
\medskip
	It should be stressed at this point that the program just outlined is strictly classical, and has for the moment no quantum version.  Note in this respect that the classical entropy may go to $-\infty$, while the quantum entropy is defined to be $\ge0$, this makes it difficult to propose a quantum nonequilibrium steady state which would be a bottomless source of entropy.
\medskip
	{\bf 11. The diversity of nonequilibrium regimes.}
\medskip
	Many important nonequilibrium systems are locally close to equilibrium, and the classical nonequilibrium studies have concentrated on that case, yielding such results as the Onsager reciprocity relations and the Green-Kubo formula.  Note however that chemical reactions are often far from equilibrium.  More exotic nonequilibrium systems of interest are provided by {\it metastable states}.  Since quantum measurements (and the associated {\it collapse of wave packets}) typically involve metastable states, one would like to have a reasonable fundamental understanding of those states.  Another class of exotic systems are {\it spin glasses}, which are almost part of equilibrium statistical mechanics, but evolve slowly, with extremely long relaxation times.
\vfill\eject

\centerline{Chapter 2. SRB STATES.}
\bigskip\bigskip
	{\bf 1. Further discussion of nonequilibrium steady states.}
\medskip
	In Chapter 1 we have proposed a definition (see (1.8)) for nonequilibrium steady states $\rho$.  We now make this definition more precise and analyze it further.  Write
$$	\rho={\rm w.lim}_{t\to\infty}\,{1\over t}
	\int_0^t d\tau\,(f^\tau)^*m\eqno{(1)}     $$
$$	m\qquad{\hbox{a.c. probability measure}}\eqno{(2)}     $$
$$	\rho\qquad{\rm ergodic}\eqno{(3)}     $$

	In (1), w.lim is the weak or vague limit defined by $({\rm w.lim}\,m_t)(\Phi)=\lim(m_t(\Phi))$ for all continuous $\Phi:M\to{\bf C}$.  The set of probability measures on $M$ is compact and metrizable for the vague topology.  There are thus always sequences $(t_k)$ tending to $\infty$ such that
$$	\rho={\rm w.lim}_{k\to\infty}\,{1\over t_k}
	\int_0^{t_k} d\tau\,(f^\tau)^*m     $$
exists; $\rho$ is automatically $(f^\tau)^*$ invariant.  By (2), we ask that the probability measure $m$ be a.c. (absolutely continuous) with respect to the Riemann volume element $dx$ (with respect to any metric) on $M$.  The condition (3) is discussed below.
\medskip
	Physically, we consider a system which in the distant past was in equilibrium with respect to a Hamiltonian time evolution
$$	{dx\over dt}=F_0(x)\eqno{(4)}     $$
and described by the probability measure $m$ on the energy surface $M$.  According to conventional wisdom, $m$ is the {\it microcanonical ensemble}, which satisfies (2), and is ergodic with respect to the time evolution (4) when the {\it ergodic hypothesis} is satisfied\footnote{*}{Even if the ergodic hypothesis is accepted, the physical justification of the microcanonical ensemble remains a delicate problem, which we shall not further discuss here.}.  For our purposes, we might also suppose that $m$ is an ergodic component of the microcanonical ensemble or an integral over such ergodic components, provided (2) is satisfied.
\medskip
	We assume now that, at some point in the distant past, (4) was replaced by the time evolution
$$	{dx\over dt}=F(x)\eqno{(5)}     $$
representing nonhamiltonian forces plus a thermostat keeping $x$ on $M$; we write the general solution of (5) as $x\mapsto f^tx$.  We are interested in time averages of $f^tx$ for $m$-almost all $x$.  Suppose therefore that
$$	\rho_x={\rm w.lim}_{t\to\infty}\,{1\over t}
	\int_0^t d\tau\,\delta_{f^\tau x}\eqno{(6)}     $$
exists for $m$-almost all $x$.  In particular, with $\rho$ defined by (1), we have
$$	\rho=\int m(dx)\,\rho_x\eqno{(7)}     $$
If (3) holds\footnote{*}{Suppose that $\rho$ is not ergodic but that $\rho=\alpha\rho'+(1-\alpha)\rho''$ with $\rho'$ ergodic and $\alpha\ne0$.  Writing $S=\{x:\rho_x=\rho'\}$, we have $m(S)=\alpha$ and $\rho'=\int m'(dx)\,\rho_x$ with $m'=\alpha^{-1}\chi_S.m$.  Therefore, (1-3) hold with $\rho$, $m$ replaced by $\rho'$, $m'$.}, (7) is equivalent to
$$	\rho_x=\rho\qquad{\hbox{for $m$-almost all $x$}}     $$
[$\Leftarrow$ is obvious; if $\Rightarrow$ did not hold (7) would give a non-trivial decomposition $\rho=\alpha\rho'+(1-\alpha)\rho''$ in contradiction with ergodicity].
\medskip
	As we have just seen, the ergodic assumption (3) allows us to replace the study of (6) by the study of (1), with the condition (2).  This has interesting consequences, as we shall see, but note that (1-3) are not always satisfyable simultaneously [consider for instance the case $F=0$].  To study nonequilibrium steady states we shall modify or strengthen the conditions (1-3) in various ways.  We may for simplicity replace the continuous time $t\in{\bf R}$ by a discrete time $t\in{\bf Z}$.  If we assume uniform hyperbolicity (Sections 2 and 3), the states satisfying (1-3) are called SRB states (Section 4).  More general SRB states have been studied by Ledrappier, Strelcyn, and Young (Section 5) and satisfy the conditions (1) under a weak (nonuniform) hyperbolicity condition.
\medskip
	{\bf 2. Uniform hyperbolicity: Anosov diffeomorphisms and flows.}
\medskip
	Let $M$ be a compact connected manifold.  In what follows we shall be concerned with a time evolution $(f^t)$ which either has discrete time $t\in{\bf Z}$, and is given by the iterates of a {\it diffeomorphism} $f$ of $M$, or has continuous time $t\in{\bf R}$, and is the {\it flow} generated by some vector field $F$ on $M$.  We assume that $f$ or $F$ are of class $C^{1+\alpha}$ (H\"older continuous first derivatives). 
\medskip
	The Anosov property is an assumption of strong chaoticity (in physical terms) or uniform hyperbolicity (in mathematical terms\footnote{**}{For background see for instance Smale [93] or Bowen [13]}).  Choose a Riemann metric on $M$.  The diffeomorphism $f$ is Anosov if there are a continuous $Tf$-invariant splitting of the tangent bundle: $TM=E^s\oplus E^u$ and constants $C>0$, $\theta>1$ such that
$$	(\forall t\ge0)\qquad\matrix{
	\|Tf^t\xi\|\le C\theta^{-t}&{\rm if}&\xi\in E^s\cr
	\|Tf^{-t}\xi\|\le C\theta^{-t}&{\rm if}&\xi\in E^u\cr}\eqno{(8)}  $$
One can show that $x\mapsto E_x^s$, $E_x^u$ are H\"older continuous, but not $C^1$ in general.  The flow $(f^t)$ is Anosov if there are a continuous $(Tf^t)$-invariant splitting\footnote{***}{Remember that $F$ is the vector field generating the flow: $F(x)=df^tx/dt|_{t=0}$.  Therefore ${\bf R}.F$ is the one-dimensional subbundle in the direction of the flow.} $TM={\bf R}.F\oplus E^s\oplus E^u$ and constants $C>0$, $\theta>1$ such that (8) again holds.
\medskip
	In what follows we shall assume that the periodic orbits are dense in $M$.  This is conjectured to hold automatically for Anosov diffeomorphisms, but there is a counterexample for flows (see [34]).  Since $M$ is connected, we are thus dealing with what is called technically a {\it mixing} Anosov diffeomorphism $f$, or a {\it transitive} Anosov flow $(f^t)$.  There is a powerful method, called {\it symbolic dynamics}, for the study of Anosov diffeomorphisms and flows.  Symbolic dynamics (see Sinai [90]) is based on the existence of {\it Markov partitions} (Sinai [91], Ratner [76]).  Another, more geometric approach based on {\it shadowing} (Bowen [13]) is also available, and will be used in Chapter 3.
\medskip
	{\bf 3. Uniform hyperbolicity: Axiom A diffeomorphisms and flows.}
\medskip
	Smale [93] has achieved an important generalization of Anosov dynamical systems by imposing hyperbolicity only on a subset $\Omega$ (the {\it nonwandering set}) of the manifold $M$.  A point $x\in M$ is {\it wandering} if there is an open set $O\ni x$ such that $O\cap f^tO\ne\emptyset$ for $|t|>1$.  The points of $M$ which are not wandering constitute the nonwandering set $\Omega$.  A diffeomorphism or flow satisfies {\it Axiom A} if the following conditions hold
\medskip
	(Aa) there is a continuous $(Tf^t)$-invariant splitting of $T_\Omega M$ (the tangent bundle restricted to the nonwandering set) verifying the hyperbolicity conditions of Section 2 above.
\medskip
	(Ab) the periodic orbits are dense in $\Omega$.
\medskip
	Under these conditions, $\Omega$ is a finite union of disjoint compact $(f^t)$-invariant sets $B$ (called {\it basic sets}) on which $(f^t)$ is topologically transitive\footnote{*}{$(f^t)$ is topologically transitive on $B$ if there is $x\in B$ such that the orbit $(f^tx)$ is dense in $B$.}.  This result is known as Smale's {\it spectral decomposition} theorem.
\medskip
	If there is an open set $U\supset B$ such that $\cap_{t\ge0}f^tU=B$, the basic set $B$ is called an {\it attractor}.  The set $\{x\in M: \lim_{t\to+\infty}d(f^tx,B)=0\}=\cup_{t\ge0}f^{-t}U$ is the {\it basin of attraction} of the attractor $B$.
\medskip
	Let $B$ be a basic set. Given $x\in B$ and $\epsilon>0$, write
$$	{\cal V}_{x,\epsilon}^s=\{y\in M: d(f^ty,f^tx)<\epsilon
{\hbox{\ for $t\ge0$, and $\lim_{t\to+\infty}d(f^ty,f^tx)=0$}}\}     $$
$$	{\cal V}_{x,\epsilon}^u=\{y\in M: d(f^ty,f^tx)<\epsilon
{\hbox{\ for $t\le0$, and $\lim_{t\to-\infty}d(f^ty,f^tx)=0$}}\}     $$
Then for sufficienty small $\epsilon$, ${\cal V}_{x,\epsilon}^s$ and ${\cal V}_{x,\epsilon}^u$ are pieces of smooth manifolds, called {\it local stable} and {\it local unstable} manifold respectively\footnote{**}{There are also {\it global} stable and unstable manifolds defined by
$$	{\cal V}_x^s=\{y\in M: \lim_{t\to+\infty}d(f^ty,f^tx)=0\}     $$
$$	{\cal V}_x^u=\{y\in M: \lim_{t\to-\infty}d(f^ty,f^tx)=0\}     $$
}, and tangent to $E_x^s$, $E_x^u$ respectively.
\medskip
	A basic set $B$ is an attractor if and only if the stable manifolds ${\cal V}_{x,\epsilon}^s$ for $x\in B$ cover a neighborhood of $B$.  Also, a basic set $B$ is an attractor if and only if the unstable manifolds ${\cal V}_{x,\epsilon}^u$ for $x\in B$ are contained in $B$ (see [14]).
\medskip
	Markov partitions, symbolic dynamics (see Bowen [9], [12]), and shadowing (see Bowen [13]) are available on Axiom A basic sets as they were for Anosov dynamical systems.  
\medskip
	{\bf 4. SRB states on Axiom A attractors} (see [92], [79], [14]).
\medskip
	Let us cut an attractor $B$ into a finite number of small cells such that the unstable manifolds are roughly parallel within a cell.  Each cell is partitionned into a continuous family of pieces of local unstable manifolds, and we obtain thus a partition $(\Sigma_\alpha)$ of $B$ into small pieces of unstable manifolds.  If $\rho$ is an invariant probability measure on $B$ (for the dynamical system $(f^t)$), and if its conditional measures $\sigma_\alpha$ with respect to the partition $(\Sigma_\alpha)$ are a.c. with respect to the Riemann volume $d\sigma$ on the unstable manifolds, $\rho$ is called an SRB measure.
\medskip
	The study of SRB measures is transformed by use of symbolic dynamics into a problem of statistical mechanics: one can characterize SRB states as Gibbs states with respect to a suitable interaction.  Such Gibbs states can in turn be characterized by a variational principle\footnote{*}{We shall come back to this topic in Chapter 3 (Section 3.4)}.  In the end one has a variational principle for SRB measures, which we shall now describe.  It is convenient to consider a general basic set $B$ (not necessarily an attractor) and to define generalized SRB measures.  First we need the concept of the {\it (time) entropy} $h_f(\mu)$, where $f=f^1$ is the time 1 map for our dynamical system, and $\mu$
an $f$-invariant probability measure on $B$.  This entropy (or Kolmogorov-Sinai invariant) has the physical meaning of mean information production per unit time by the dynamical system $(f^t)$ in the state $\mu$ (see for instance [6] for an exact definition).  The time entropy $h_f(\mu)$ should not be confused with the {\it space} entropy $S$ and the entropy production rate $e_f$ which we have discussed in Chapter 1.  We also need the concept of {\it expanding Jacobian} $J^u$.  Since $T_xf$ maps $E_x^u$ linearly to $E_{fx}^u$, and volume elements are defined in $E_x^u$, $E_{fx}^u$ by  a Riemann metric, we can define a volume expansion rate $J^u(x)>0$.  It can be shown that the function $\log J^u:B\to{\bf R}$ is H\"older continuous.  We say that the $(f^t)$-invariant probability measure $\rho$ is a {\it generalized SRB measure} if it makes maximum the function
$$	\mu\quad\mapsto\quad h_f(\mu)-\mu(\log J^u)\eqno{(9)}     $$
One can show that there is precisely one generalized SRB measure on each basic set $B$; it is ergodic and reduces to the unique SRB measure when $B$ is an attractor.  The value of the maximum of (9) is 0 precisely when $B$ is an attractor, it is $<0$ otherwise.
\medskip
	If $m$ is a measure absolutely continuous with respect to the Riemann volume element on $M$, and if its density $\underline m$ vanishes outside of the basin of attraction of an attractor $B$, then
$$	\rho={\rm w.lim}_{t\to\infty}\,{1\over t}
	\int_0^t d\tau\,(f^\tau)^*m\eqno{(10)}     $$
defines the unique SRB measure on $B$.  We also have
$$	\rho={\rm w.lim}_{t\to\infty}\,{1\over t}
	\int_0^t d\tau\,\delta_{f^\tau x}     $$
when $x$ is in the domain of attraction of $B$ and outside of a set of measure 0 for the Riemann volume.  The conditions (1-3) and also (6) are thus satisfied, and the SRB state $\rho$ is a natural nonequilibrium steady state.  Note that if $B$ is a {\it mixing}\footnote{*}{i.e., for any two nonempty open sets $O_1$, $O_2\subset B$, there is $T$ such that $O_1\cap f^tO_2\ne\emptyset$ for $|t|\ge T$.} attractor, (10) can be replaced by the stronger result
$$	\rho={\rm w.lim}_{t\to\infty}\,(f^t)^*m     $$
\medskip
	{\bf 5. SRB states without uniform hyperbolicity} (see [66], [67], [95]).
\medskip
	Remarkably, the theory of SRB states on Axiom A attractors extends to much more general situations.  Consider a smooth dynamical system $(f^t)$ on the compact manifold $M$, without any hyperbolicity condition, and let $\rho$ be an ergodic measure for this system.  The theorem of Oseledec [69], [82] permits the definition of {\it Lyapunov exponents} $\lambda_1\le\ldots\le\lambda_{{\rm dim}\, M}$ which are the rates of expansion, $\rho$-almost everywhere, of vectors in $TM$.  The $\lambda_i$ are real numbers, positive (expansion), negative (contraction), or zero (neutral case).  Pesin theory [71], [72] allows the definition of stable and unstable manifolds $\rho$-almost everywhere.  These are smooth manifolds; the dimension of the stable manifolds is the number of negative Lyapunov exponents while the dimension of the unstable manifold is the number of positive Lyapunov exponents.  Consider now a family $(\Sigma_\alpha)$ constituted of pieces of (local) unstable manifolds, and forming, up to a set of $\rho$-measure 0, a partition of $M$.  As in Section 4 above we define the conditional measures $\sigma_\alpha$ of $\rho$ with respect to $(\Sigma_\alpha)$.  If the measures $\sigma_\alpha$ are absolutely continuous with respect to the Riemann volumes of the corresponding unstable manifolds, we say that $\rho$ is an SRB measure.
\medskip
	For a $C^2$ diffeomorphism, the above definition of SRB measures is equivalent to the following condition (known as {\it Pesin formula}) for an ergodic measure $\rho$:
$$	h(\rho)={\hbox{sum of the positive
	Lyapunov exponents for $\rho$}}     $$
(see [66], [67] for the nontrivial proof).  This is an extension of the result of Section 4, where the sum of the positive Lyapunov exponents is equal to $\rho(\log J^u)$.  Note that in general $h(\mu)$ is $\le$ the sum of the positive exponents for the ergodic measure $\mu$ (see [80]).
\medskip
	Suppose that the time $t$ is discrete (diffeomorphism case), and that the Lyapunov exponents of the SRB state $\rho$ are all different from zero: this is a weak (nonuniform) hyperbolicity condition.  In this situation, there is a measurable set $S\subset M$ with positive Riemann volume such that
$$  \lim_{n\to\infty}{1\over n}\sum_{k=0}^{n-1}\delta_{f^kx}=\rho  $$
for all $x\in S$.  This result [75] shows that $\rho$ has the properties required of a nonequilibrium steady state.  One expects that for continuous time (flow case), if supp $\rho$ is not reduced to a point and if the Lyapunov exponents of the SRB state $\rho$ except one\footnote{*}{There is one zero exponent corresponding to the flow direction.} are different from 0, there is a measurable set $S\subset M$ with positive Riemann volume such that
$$	\rho={\rm w.lim}_{t\to\infty}\,{1\over t}
	\int_0^t d\tau\,\delta_{f^\tau x}     $$
when $x\in S$.  I do not know a proof for this, however.
\medskip
	After Pesin's theory, a more geometric approach to dynamical systems without uniform hyperbolicity has started to develop (work of Jakobson, Benedicks, Carleson, Young).  This is a difficult topic, which will not be dealt with here, but we refer the reader to the excellent review of L.-S. Young [99].  The interest of the geometric approach is that it gives more detailed results than the general Pesin theory: the existence of an SRB measure is proved instead of being assumed and the (usually exponential) decay of correlations can be studied, see Young [97], [98].  There is currently a rapidly growing literature on this topic, see for instance Viana [95], Chernov [15].
\medskip
	{\bf 6. Rigorous versus formal arguments.}
\medskip
	We have just outlined some possible mathematical descriptions of nonequilibrium steady states.  In the present situation however, where the theory is still in its infancy, priority should be given to getting the physics right.  For the mathematical treatment, two main attitudes appear reasonable:
\medskip
	(a) We may make uniform hyperbolicity assumptions -- that the time evolution is given by an Anosov or Axiom A diffeomorphism or flow -- and proceed rigorously.  Of course uniform hyperbolicity is quite unrealistic from a physical point of view.
\medskip
	(b) We may proceed formally and assume when needed that limits exist, that Lyapunov exponents are nonzero, or that correlation functions decrease fast.
\medskip
	There are several reasons not to press too hard now the mathematical study of nonequilibrium steady states in the general situation of Section 5.  One is that the mathematics of nonuniformly hyperbolic systems is still under development (for an important recent result see [5]).  Another is that we shall, for physical reasons, want to take derivatives of SRB states with respect to parameters (Chapter 4), while these derivatives appear to make no mathematical sense in general.  The resolution of this difficulty may involve the need to take a large system limit (thermodynamic limit) but this remains to be investigated.
\vfill\eject

\centerline{Chapter 3. THE GALLAVOTTI-COHEN FLUCTUATION THOREM.}
\bigskip\bigskip
	{\bf 1. Generalities.}
\medskip
	In this Chapter we prove an asymptotic formula due to Gallavotti and Cohen [44], [45] for fluctuations of the entropy production .  The success of this formula supports the philosophy that calculations based on very strong hyperbolicity assumptions are in agreement with experiments (in this case numerical experiments).  Following Gallavotti and Cohen we take the time to be discrete ($t\in{\bf Z}$), and the evolution to be given by an Anosov diffeomorphism (see Section 2.2) on the compact connected manifold $M$.  We shall follow the ideas of Gallavotti and Cohen but change slightly the presentation.  Instead of {\it Markov partitions} and {\it symbolic dynamics} we shall use a more geometric approach based on {\it shadowing} (Bowen [13]).
\medskip
	{\bf 2. Expansiveness, $\alpha$-pseudoorbits, shadowing, and specification for Anosov diffeomorphisms.}
\medskip
	An Anosov diffeomorphism is {\it expansive}, {\it i.e.}, there is $\delta>0$ (called {\it expansive constant}) such that
$$	((\forall k\in{\bf Z})\enspace d(f^kx,f^ky)<\delta)
	\qquad\Longleftrightarrow\qquad x=y     $$
[In view of (2.8), when two orbits are close to each other, they must separate exponentially in the past or in the future, unless they coincide].  In fact we have, for some $K>0$, $\theta>1$,
$$(d(f^kx,f^ky)<\delta\enspace{\rm when}\enspace|k|\le n)\qquad\Rightarrow
	\qquad d(x,y)<K\theta^{-n}\eqno{(1)}     $$
\medskip
	Let $S$ be an interval of ${\bf Z}$, finite or not, and let ${\bf x}=(x_k)_{k\in S}\in M^S$.  Given $\alpha>0$, we say that ${\bf x}$ is an $\alpha$-{\it pseudoorbit} if
$$	d(fx_k,x_{k+1})<\alpha\qquad{\rm whenever}\qquad k,k+1\in S     $$
We say that the orbit of $x\in M$ $\alpha$-{\it shadows} ${\bf x}$ if
$$	d(f^kx,x_k)<\alpha\qquad{\rm for}\enspace{\rm all}\qquad k\in S     $$
\par
	{\bf Lemma.} (Bowen's shadowing lemma [13]).
\medskip
	{\sl Given $\beta>0$ there is $\alpha>0$ such that every $\alpha$-pseudoorbit is $\beta$-shadowed by the orbit of a point $x\in M$.}
\medskip
	Note that if the pseudoorbit is periodic, then the true orbit which shadows it over ${\bf Z}$ is unique by (1), and periodic with the same period.  [We assume that $2\beta$ is an expansive constant].
\medskip
	Given $\alpha>0$ there is an integer $R>0$ such that for every pair $(x,y)$ of points of $M$, and $n\ge R$, there is $z$ such that
$$	d(z,x)<\alpha\qquad,\qquad d(f^nz,y)<\alpha     $$
This follows from the fact that $f$ is topologically mixing (because we assumed $f$ Anosov and $M$ connected, see Section 2.2), and says that there is an orbit which interpolates in time $n$ between any two points $x$ and $y$.  Using the shadowing lemma, we see that there is a true orbit interpolating between any number of $\alpha$-pseudoorbits specified on on intervals with distances $\ge R$.  This interpolation property is known as {\it specification} (see Bowen [10]).
\medskip
	{\bf Lemma.} (Specification property).
\medskip
	{\sl Let $\alpha$-pseudoorbits ${\bf x}_j$ be given on disjoint intervals of ${\bf Z}$ with mutual distances at least $R$.  Then the ${\bf x}_j$ are all $\beta$-shadowed by the orbit of the same point $x\in M$.}
\medskip
	{\bf 3. Approximation of sums along orbits.}
\medskip
	We shall later have to consider sums of the form
$$	\sum_{k=a}^bA(f^kx)     $$
If $f^ky$ stays close to $f^kx$ for $a\le k\le b$, we know by (1) that $d(f^kx,f^ky)\le K\theta^{-n}$ where $n=\min(k-a,b-k)$.  Therefore, if $A$ is H\"older continuous\footnote{*}{Remember that $A$ is $\gamma$-H\"older continuous (with $0<\gamma\le1$) if $|A(x)-A(y)|<Cd(x,y)^\gamma$.  Thus $|A(f^kx)-A(f^ky)|\le Cd(f^kx,f^ky)^\gamma\le CK^\gamma(\theta^\gamma)^{-n}$.}
$$	|\sum_{k=a}^bA(f^kx)-\sum_{k=a}^bA(f^ky)|\le L     $$
where $L$ depends only on $A$.
\medskip
	Suppose now that we use specification to obtain from $x$ and $y$ a periodic point $z$ with $f^nz=z$ and
$$	f^kz\qquad{\rm close\enspace to}\qquad f^kx\qquad{\rm for}
	\qquad k=0,\ldots,\tau-1     $$
$$	f^kz\qquad{\rm close\enspace to}\qquad f^ky\qquad{\rm for}
	\qquad k=\tau+R-1,\ldots,n-R     $$
then we also have
$$	|\sum_{k=0}^{n-1}A(f^kz)-\sum_{k=0}^{\tau-1}A(f^kx)
	-\sum_{k=\tau+R-1}^{n-R}A(f^ky)|\le L'\eqno{(2)}     $$
where $L'$ depends only on $A$.
\medskip
	{\bf 4. Periodic orbit representation of Gibbs states and SRB states for Anosov diffeomorphisms.}
\medskip
	We assume from now on that $f$ is (at least) of class $C^{1+\alpha}$ for some $\alpha>0$.  If $A:M\to{\bf R}$ is H\"older continuous it is known that there is a unique $f$-ergodic measure $\rho$ making
$$	h(\rho)+\int A(x)\rho(dx)     $$
maximum; $\rho$ is called the {\it equilibrium state} or {\it Gibbs state for $A$} (see Ruelle [81]).
\medskip
	We need at this point the {\it unstable Jacobian} $J^u$ (see Section 2.4).  By definition $(Tf)^{\wedge{\rm dim}E^u}$ maps the volume element of $E_x^u$ to $J^u(x)$ times the volume element of $E_{fx}^u$, where the volume elements are defined with respect to the Riemann metric, and $J^u(x)>0$.  Our assumption that $f$ is $C^{1+\alpha}$ implies that $J^u$ is H\"older continuous.  In the present situation of an Anosov diffeomorphism, the SRB state was shown by Sinai [92] to be the Gibbs state $\rho_f$ for $A=-\log J^u$.  Note that we have (for historical reasons) a somewhat confusing terminology: {\it nonequilibrium} steady states for a certain problem are described by {\it equilibrium} states (or Gibbs states) for another (very different) problem.  There is however no ambiguity when we speak of SRB states.
\medskip
	Let us write ${\rm Fix}f^n=\{x\in M:f^nx=x\}$.  An f-invariant probability measure ${\bf\sigma}_n$, carried by periodic orbits, is defined by
$$	{\bf\sigma}_n(\Phi)=
	{\sum_{x\in{\rm Fix}f^n}\Phi(x)\exp\sum_{k=0}^{n-1}A(f^kx)
	\over\sum_{x\in{\rm Fix}f^n}\exp\sum_{k=0}^{n-1}A(f^kx)}     $$
and it is known that, when $n\to\infty$, ${\bf\sigma}_n$ tends weakly to the Gibbs state for $A$ ({\it i.e.}, ${\bf\sigma}_n(\Phi)$ tends to $\rho(\Phi)$ for all continuous $\Phi$, see [81]).  In particular, the SRB state $\rho_f$ satisfies
$$	\rho_f(\Phi)=\lim_{n\to\infty}
	{\sum_{x\in{\rm Fix}f^n}\Phi(x)\prod_{k=0}^{n-1}J^u(f^kx)^{-1}
\over\sum_{x\in{\rm Fix}f^n}\prod_{k=0}^{n-1}J^u(f^kx)^{-1}}\eqno{(3)}     $$
\medskip
	{\bf 5. Reversibility.}
\medskip
	We say that the time evolution $(f^t)$, with $t\in{\bf Z}$, or ${\bf R}$ is {\it reversible} if there is a diffeomorphism $i:M\to M$ such that $i\circ i={\rm identity}$ and $i\circ f^t=f^{-t}\circ i$.  [Note that any $i\circ f^s$ then satisfies the same conditions: $i\circ f^s\circ i\circ f^s=i\circ i\circ f^{-s}\circ f^s={\rm identity}$, and $i\circ f^s\circ f=i\circ f^{s+1}=f^{-1}\circ i\circ f^s$].  We choose, as we may, a Riemann metric on $M$ invariant under $i$.  This gives in particular
$$	J(ifx)=J(x)^{-1}\eqno{(4)}     $$
and, since $i$ interchanges the stable and unstable directions, also
$$	J^s(ifx)=J^u(x)^{-1}\eqno{(5)}      $$
For a system of particles with positions $x_k$ and momenta $p_k$ one may take $i((x_k),(p_k))=((x_k),(-p_k))$ if the forces are not velocity dependent (electrostatic forces are permitted, but not magnetic fields).  In this physical situation one speaks of {\it microscopic reversibility}.  This was an essential ingredient in Onsager's proof of his reciprocity relations, and it is also essential for the fluctuation theorem of Gallavotti and Cohen.
\medskip
	{\bf 6. Entropy production.}
\medskip
	We have discussed earlier (Section 1.9) the formula
$$	e_f=-\int\rho_f(dx)\log J(x)\ge0     $$
for the average entropy production per unit time in the nonequilibrium steady state defined by the SRB measure $\rho_f$.  Remember that $J>0$ is the Jacobian of $f$ with respect to the chosen metric on $M$.  In the situation which interests us here, where $f$ is an Anosov diffeomorphism of the connected manifold $M$, we have $e_f>0$ unless $f$ has an invariant measure absolutely continuous with respect to the Riemann volume.  [This follows from Ledrappier [65], see the discussion in [85]].
\medskip
	For integer $\tau>0$, let
$$	\epsilon_\tau(x)
={1\over\tau e_f}\sum_{k=0}^{\tau-1}\log J(f^kx)^{-1}\eqno{(6)}     $$
This may be interpreted as a dimensionless entropy production rate or phase space contraction rate at $x$ over time $\tau$, such that $\rho_f(\epsilon_\tau)=1$.  Note that if we have reversibility, (4) yields
$$	\epsilon_\tau\circ i\circ f^\tau=-\epsilon_\tau\eqno{(7)}     $$
The fluctuation theorem of Gallavotti and Cohen (Theorem 9 below) is a statement about the fluctuations of $\epsilon_\tau$ with respect to the SRB measure $\rho_f$ (or equivalently with respect to time).
\medskip
	{\bf 7. Proposition.}
\medskip
	{\sl Let $f$ be a $C^{1+\alpha}$ Anosov diffeomorphism of the compact connected manifold $M$, and let $\rho_f$ be the corresponding SRB measure.  Assume reversibility ($i\circ f=f^{-1}\circ i$, $i^2={\rm identity}$) and define $\epsilon_\tau$ by (6) with respect to an $i$-invariant Riemann metric on $M$.  Under these conditions there are constants $a$, $b>0$ such that for any interval $[p,q]\subset{\bf R}$,
$$	{1\over\tau e_f}\log{\rho_f(\{x:\epsilon_\tau(x)\in[p,q]\})
	\over\rho_f(\{x:\epsilon_\tau(x)\in[-q-a/\tau,-p+a/\tau]\})}
	\le q+{b\over\tau}\eqno{(8)}     $$}
\par
	Let $\chi$ be the characteristic function of $[p,q]$ and $\chi^*$ the characteristic function of $[p-a/\tau,q+a/\tau]$ where $a$ will be fixed later.  Using (7) we have
$$	{\rho_f(\{x:\epsilon_\tau(x)\in[p,q]\} 
	\over\rho_f(\{x:\epsilon_\tau(x)\in[-q-a/\tau,-p+a/\tau]\}}
={\rho_f(\chi\circ\epsilon_\tau)\over\rho_f(\chi^*\circ(-\epsilon_\tau))}    $$
$$={\rho_f(\chi\circ\epsilon_\tau)\over\rho_f(\chi^*\circ\epsilon_\tau\circ i)}
     ={\rho_+(\chi\circ\epsilon_\tau)\over\rho_-(\chi^*\circ\epsilon_\tau)}  $$
where $\rho_+=\rho_f$ and $\rho_-=i^*\rho_f$ is the anti-SRB state, {\it i.e.}, the Gibbs state for $\log J^s$ (where $J^s$ is the {\it stable Jacobian}).  We replace now $\chi$, $\chi^*$ by continuous functions $\phi$, $\phi^*$ (we shall take $\phi\downarrow\chi$, $\phi^*\downarrow\chi^*$ later) and use the formula (3) (approximation of SRB states by periodic orbits) to rewrite
$$   {\rho_+(\phi\circ\epsilon_\tau)\over\rho_-(\phi^*\circ\epsilon_\tau)}   $$
$$	=\lim_{n\to\infty}
{\sum_{x\in{\rm Fix}f^n}\phi(\epsilon_\tau(x))\prod_{k=0}^{n-1}J^u(f^kx)^{-1}
\over\sum_{x\in{\rm Fix}f^n}\prod_{k=0}^{n-1}J^u(f^kx)^{-1}}:
{\sum_{x\in{\rm Fix}f^n}\phi^*(\epsilon_\tau(x))\prod_{k=0}^{n-1}J^s(f^kx)
\over\sum_{x\in{\rm Fix}f^n}\prod_{k=0}^{n-1}J^s(f^kx)}     $$
$$	=\lim_{n\to\infty}
{\sum_{x\in{\rm Fix}f^n}\phi(\epsilon_\tau(x))\prod_{k=0}^{n-1}J^u(f^kx)^{-1}
	\over\sum_{x\in{\rm Fix}f^n}\phi^*(\epsilon_\tau(x))
	\prod_{k=0}^{n-1}J^s(f^kx)}     $$
[The last step is based on (5) and the fact that $i$ is a bijection ${\rm Fix}f^n\to{\rm Fix}f^n$ which maps orbits of $f$ to orbits of $f$ (reversing the direction)].
\medskip
	If $x\in {\rm Fix}f^n$, let $S(x)\subset{\rm Fix}f^n$ consist of those $y$ such that $f^k y$ is close to $f^kx$ for
$k=0,\ldots,\tau-1$, and to $if^{n-k+\tau}x$ for $k=\tau+R-1,\ldots,n-R$.  By specification ${\rm card}S(x)\ge1$, and by expansiveness ${\rm card}S(x)\le N$ for some $N$ independent of $n$, $\tau$.  Suppose $y\in S(x)$, in view of (2) and (5) we have
$$	\prod_{k=0}^{n-1}J^u(f^ky)^{-1}\le M
   \prod_{k=0}^{\tau-1}J^u(f^kx)^{-1}\prod_{k=\tau}^{n-1}J^s(f^kx)   $$
with $M$ independent of $n$, $\tau$.  We also have 
$$	|\epsilon_\tau(y)-\epsilon_\tau(x)|<{a\over\tau}     $$
with $a$ independent of $n$, $\tau$ (this fixes our choice of $a$).  Therefore $\chi(\epsilon_\tau(y))\le\chi^*(\epsilon_\tau(x))$, and we may assume that the continuous functions $\phi$, $\phi^*$ (such that $\phi\downarrow\chi$, $\phi^*\downarrow\chi^*$) also satisfy $\phi(\epsilon_\tau(y))\le\phi^*(\epsilon_\tau(x))$.  Thus
$$	\sum_{x\in{\rm Fix}f^n}\phi(\epsilon_\tau(x))
	\prod_{k=0}^{n-1}J^u(f^kx)^{-1}      $$
$$	\le\sum_{x\in{\rm Fix}f^n}\sum_{y\in S(x)}\phi(\epsilon_\tau(y))
	\prod_{k=0}^{n-1}J^u(f^ky)^{-1}      $$
$$	\le\sum_{x\in{\rm Fix}f^n}{\rm card}S(x)\phi^*(\epsilon_\tau(x))
	M\prod_{k=0}^{\tau-1}J^u(f^kx)^{-1}\prod_{k=\tau}^{n-1}J^s(f^kx)    $$
$$	\le MN\sum_{x\in{\rm Fix}f^n}\phi^*(\epsilon_\tau(x))
	\prod_{k=0}^{\tau-1}J^u(f^kx)^{-1}\prod_{k=\tau}^{n-1}J^s(f^kx)    $$
Hence
$$   {\rho_+(\phi\circ\epsilon_\tau)\over\rho_-(\phi^*\circ\epsilon_\tau)}\le
	\lim_{n\to\infty}MN{\sum_{x\in{\rm Fix}f^n}\phi^*(\epsilon_\tau(x))
	\prod_{k=0}^{\tau-1}J^u(f^kx)^{-1}\prod_{k=\tau}^{n-1}J^s(f^kx)
	\over\sum_{x\in{\rm Fix}f^n}\phi^*(\epsilon_\tau(x))
\prod_{k=0}^{\tau-1}J^s(f^kx)\prod_{k=\tau}^{n-1}J^s(f^kx)}\eqno{(9)}     $$
We still need the fact that
$$	{\prod_{k=0}^{\tau-1}J^s(f^kx)\prod_{k=0}^{\tau-1}J^u(f^kx)
	\over\prod_{k=0}^{\tau-1}J(f^kx)}\ge P     $$
where the constant $P>0$ is independent of $x$ and $\tau$ (this follows from the continuity of the splitting $TM=E^s\oplus E^u$, and compactness).  Thus, by (6),
$$	{\prod_{k=0}^{\tau-1}J^u(f^kx)^{-1}\over\prod_{k=0}^{\tau-1}J^s(f^kx)}
	\le{1\over P}\exp(\tau e_f\epsilon_\tau(x))     $$
Inserting in (9) and performing the limit $n\to\infty$ we find, using Section 4 and the definition of $\rho_-$,
$$   {\rho_+(\phi\circ\epsilon_\tau)\over\rho_-(\phi^*\circ\epsilon_\tau)}\le
{MN\over P}{\rho_-((\phi^*\circ\epsilon_\tau).\exp(\tau e_f\epsilon_\tau))
\over\rho_-(\phi^*\circ\epsilon_\tau)}      $$
Letting now $\phi^*\downarrow\chi^*$ and using $\epsilon_\tau(x)\le q+a/\tau$ when $\chi^*(\epsilon_\tau(x))\ne0$, we obtain
$$   {\rho_+(\chi\circ\epsilon_\tau)\over\rho_-(\chi^*\circ\epsilon_\tau)}\le
	{MN\over P}\exp(\tau e_f(q+{a\over\tau}))     $$
which yields (8) if we take $b=a+(1/e_f)\log(MN/P)$.\qed
\medskip
	{\bf 8. A large deviation result.}
\medskip
	Proposition 7 is a first form of the fluctuation theorem.  To obtain the final form we need the following {\it large deviation} result.
\medskip
	{\sl Let $A$, $B:M\to{\bf R}$ be H\"older continuous, and let $\rho$ be a Gibbs state for $A$.  There exists then a real concave function $\eta$ on the open interval $(p_1^*,p_2^*)$ such that, for every interval $I$ with $I\cap(p_1^*,p_2^*)\ne0$,
$$	\lim_{\tau\to\infty}{1\over\tau}\log\rho(\{x:{1\over\tau}
	\sum_{k=0}^{\tau-1}B(f^kx)\in I\})
	=\sup_{u\in I\cap(p_1^*,p_2^*)}\eta(u)\eqno{(10)}     $$
[We disregard here a degenerate case where $p_1^*=p_2^*$ and $(1/\tau)\sum_{k=0}^{\tau-1}B(f^kx)$ does not fluctuate for large $\tau$].}
\par
	By use of symbolic dynamics, this is converted into a fact of equilibrium statistical mechanics, for which see Lanford [59].  Because we have here one-dimensional statistical mechanics with short range interactions, $\eta$ is in fact real analytic and strictly concave.
\medskip
	We take now $A=-\log J^u$ so that $\rho=\rho_f$, and we let $B=-(1/e_f)\log J$.  We have thus $(1/\tau)\sum_{k=0}^{\tau-1}B(f^kx)=\epsilon_\tau(x)$.  Also, we have $(p_1^*,p_2^*)=(-p^*,p^*)$, {\it i.e.}, $\epsilon_\tau$ fluctuates in an interval symmetric around $0$; this follows from reversibility as should become clear in a moment.  Equation (10) now has the form
$$	\lim_{\tau\to\infty}{1\over\tau e_f}\log
	\rho_f(\{x:\epsilon_\tau(x)\in I\})
	=\sup_{u\in I\cap(-p^*,p^*)}\tilde\eta(u)     $$
(with $\tilde\eta=\eta/e_f$).  In particular, if $|p|<p^*$ and $\delta>0$ we have
$$	\lim_{\tau\to\infty}{1\over\tau e_f}\log
	\rho_f(\{x:\epsilon_\tau(x)\in(-p-\delta,-p+\delta)\})     $$
$$	=\lim_{\tau\to\infty}{1\over\tau e_f}\log\rho_f
(\{x:\epsilon_\tau(x)\in(-p-\delta-{a\over\tau},-p+\delta+{a\over\tau})\})$$
From (8) we obtain also
$$	\lim_{\tau\to\infty}{1\over\tau e_f}\log{
	\rho_f(\{x:\epsilon_\tau(x)\in(p-\delta,p+\delta)\})\over\rho_f
(\{x:\epsilon_\tau(x)\in(-p-\delta-a/\tau,-p+\delta+a/\tau)\})}
	\le p+\delta     $$
Therefore
$$	\lim_{\tau\to\infty}{1\over\tau e_f}\log{
	\rho_f(\{x:\epsilon_\tau(x)\in(p-\delta,p+\delta)\})\over\rho_f
	(\{x:\epsilon_\tau(x)\in(-p-\delta,-p+\delta)\})}
	\le p+\delta\eqno{(11)}     $$
\par
	{\bf 9. Theorem} (Gallavotti-Cohen fluctuation theorem [44], [45], [36]).
\medskip
	{\sl Let $f$ be a $C^{1+\alpha}$ Anosov diffeomorphism of the compact connected manifold $M$, and let $\rho_f$ be the corresponding SRB measure.  Assume reversibility ($i\circ f=f^{-1}\circ i$, $i^2={\rm identity}$) and define $\epsilon_\tau$ by (6) with respect to an $i$-invariant Riemann metric on $M$.  Then there exists $p^*>0$ such
that if $|p|<p^*$ and $\delta>0$
$$	p-\delta\le\lim_{\tau\to\infty}{1\over\tau e_f}\log{
	\rho_f(\{x:\epsilon_\tau(x)\in(p-\delta,p+\delta)\})\over\rho_f
	(\{x:\epsilon_\tau(x)\in(-p-\delta,-p+\delta)\})}\le p+\delta $$}
\medskip
	This follows from (11) and the formula
$$	\lim_{\tau\to\infty}{1\over\tau e_f}\log{
	\rho_f(\{x:\epsilon_\tau(x)\in(-p-\delta,-p+\delta)\})\over\rho_f
	(\{x:\epsilon_\tau(x)\in(p-\delta,p+\delta)\})}\le -p+\delta $$
obtained by changing $p$ to $-p$ in (11).\qed
\medskip
	{\bf 10. Extension of the fluctuation theorem to time-reversed events.}
\medskip
	Inspection of the proof of the Gallavotti-Cohen fluctuation theorem shows that, basically, it compares the probability $P_+$ of the event $\epsilon_\tau(x)\approx p$ with the probability $P_-$ of the time reversed event $\epsilon_\tau(if^\tau x)\approx p$, obtaining $P_+/P_-\approx e^{e_fp\tau}$.  As remarked by Gallavotti [43], one might as well compare (for any function $\Psi$) the probability $\tilde P_+$ of the event
$$	\epsilon_\tau(x)\approx p\qquad{\rm and}\qquad
	\Psi(x)\approx A     $$
with the probability $\tilde P_-$ of the event
$$	\epsilon_\tau(if^\tau x)\approx p\qquad{\rm and}\qquad
	\Psi(if^\tau x)\approx A     $$
again with the result $\tilde P_+/\tilde P_-\approx e^{e_fp\tau}$.  Note that $\Psi(x)\approx A$ could for instance describe density fluctuations of a system of particles in the time interval $[0,\tau]$, and $\Psi\circ i\circ f^\tau(x)\approx A$ the time reversed fluctuations in the same interval $[0,\tau]$.  Also note that the right hand side of $\tilde P_+/\tilde P_-\approx e^{e_fp\tau}$ depends on $p$ but not on $A$.  A specific example of an extension of the Gallavotti-Cohen fluctuation theorem is the following formula, where $\Psi$ is a strictly positive continuous function on $M$:
$$	p-\delta\le\lim_{\tau\to\infty}{1\over\tau e_f}\log{
	\int_{|\epsilon_\tau-p|<\delta}\rho_f(dx)\Psi(x)\over
	\int_{|\epsilon_\tau+p|<\delta}\rho_f(dx)\Psi(i f^\tau x)}
	\le p+\delta     $$
\par
	Remember that $\rho_f(\epsilon_\tau)=1$, so that $\epsilon_\tau$ takes mostly positive values.  Gallavotti [43] points out that in those exceptional cases when $\epsilon_\tau(x)\approx -p<0$ one observes in the interval $[0,\tau-1]$ (roughly speaking) the time reversed behavior from the "normal" behavior corresponding to $\epsilon_\tau(x)\approx p<0$.  This means that the pattern of time fluctuations is reversed, {\it i.e.}, the time appears to flow backwards.
\medskip
	{\bf 11. Other extensions.}
\medskip
	Note that the results of this chapter apply {\it far from equilibrium} (we assume that the time evolution is reversible, but we do not assume that it is close to Hamiltonian).  Placing ourselves {\it close to equilibrium} we could obtain from the fluctuation theorem a linear response formula called {\it fluctuation-dissipation theorem}, and the {\it Onsager reciprocity relations} (see Gallavotti [37]).  We prefer to postpone the derivation of these results, and obtain in the next chapter a linear response formula far from equilibrium and without assumption of reversibility.
\medskip
	Naturally, one wants to extend the fluctuation theorem from Anosov diffeomorphisms to Anosov flows.  Such an extension has been obtained by Gentile [49],using Markov partitions.  Probably one could use instead an approximation of SRB states by periodic orbits, and Bowen's techniques (see [11] and [70]).
\medskip
	Physically, the assumption of an Anosov time evolution is not as unrealistic as it would at first sight appear.  In fact, close to equilibrium, hard sphere systems are expected to be hyperbolic (and differ from Anosov systems only by the presence of singularities).  Problems arise, however, as we move away from equilibrium.  In particular, time reversibility implies that the stable and unstable dimensions (${\rm dim}E^s$ and ${\rm dim}E^u$) are equal to ${\rm dim}M/2$ for an Anosov diffeomorphism (or $({\rm dim}M-1)/2$ for an Anosov flow).  Therefore the numbers of positive and negative Lyapunov exponents are also equal (for any ergodic measure, in particular the SRB measure $\rho_f$).  If pairing is satisfied (see Dettmann-Morriss [22]) the Lyapunov exponents come by pairs $\lambda_{-i}$, $\lambda_i$ with $\lambda_{-i}+\lambda_i=2a$ independent of $i$ (for the SRB measure $\rho_f$, we have $a=-e_f/{\rm dim}M$, resp. $a=-e_f/({\rm dim}M-1)$ for a diffeomorphism resp flow).  The Anosov assumption thus forces a gap of size $2|a|$ between the positive and the negative exponents, which seems unphysical.  To avoid this problem, Bonetto and Gallavotti [7] consider a situation where $f$ satisfies the Axiom A of Smale [93], but the SRB measure $\rho_f$ and the anti-SRB measure $i^*\rho_f$ may now have disjoint supports.  Assuming a new symmetry (obtained from what is called Axiom C in [7]), Bonetto and Gallavotti can still derive the fluctuation theorem.  In the absence of uniform hyperbolicity, the number of positive Lyapunov exponents may vary discontinuously, and it is difficult to relate $\rho_f$ and $i^*\rho_f$ even if they have the same support; a fluctuation theorem has not been derived under these more general circumstances.
\medskip
	The available numerical simulations (by Evans, Cohen, Morriss
[31], and Bonetto, Gallavotti and Garrido [8]) are in excellent
agreement with the fluctuation theorem, but in a situation where the
equality between the numbers of positive and negative Lyapunov exponents
is respected.  Note that, in the form given above, the fluctuation
theorem applies to {\it small} systems; for large systems, the entropy
production fluctuates very little, and negative values are unobservably
rare.  Following a suggestion by Gallavotti, there is currently a lot of
interest in a possible {\it local} fluctuation theorem (which would
apply to small subsystems of a large system).  For other interesting
recent work, see Kurchan [58], Lebowitz and Spohn [64].
\vfill\eject
\centerline{Chapter 4. LINEAR RESPONSE.}
\bigskip\bigskip
	{\bf 1. Setting the problem.}
\medskip
	We shall see that the definition of SRB state extends to time dependent dynamics (nonautonomous systems).  This gives in particular the possibility of studying the response of a stationary nonequilibrium state to small time dependent perturbations.  The linear response function thus defined is a physically important quantity, and its study is the main subject of this chapter.  Our approach to computing the linear response will be mostly formal.  Actually, in the uniformly hyperbolic case, one can rigorously establish the differentiability implied by linear response, but it should be realized that this situation is exceptional.  In general, {\it i.e.}, unless we are in some kind of structurally stable situation, changing the parameters of a dynamical system (unless the changes are in special directions) will entail rapid discontinuous changes in the dynamics, and also in the SRB states.  From a physical point of view this may not be as bad as it seems, because in physical applications we are mostly interested in large systems.  For such systems ({\it i.e.}, in the {\it thermodynamic limit}) one can hope that the dependence on parameters is in general smooth, and that our linear response formula is physically correct.  To put things more bluntly, if there is a linear response formula at the level of experimentally observable quantities, it must be (10) as discussed below.  This is because our derivation, altough formal, offers no reasonable alternative.  What may happen is that the integral in the right-hand side of (10) diverges, leaving us without a response formula.
\medskip
	The discussion of the present chapter is valid {\it far from equilibrium}, {\it i.e.}, valid in general.  The assumption of closeness to equilibrium and its consequences will be considered in the next chapter.  While our derivation of the linear response formula will be formal, and easy, it involves a nontrivial element, {\it viz.}, the SRB state $\rho$.  By contrast with the situation close to equilibrium, where one can make guesses on the basis of general physical principles, we shall in what follows need specific properties of SRB measures.
\medskip
	{\bf 2. Time dependent SRB states.}
\medskip
	It will be convenient to assume for the purposes of this section that we have a discrete time $t\in{\bf Z}$.  Instead of a single diffeomorphism $f$ of the compact manifold $M$ we consider a family $(f_t)_{t\in{\bf Z}}$, {\it i.e.}, we have time dependent dynamics.  What we want to call SRB state in this situation is a family $(\rho_t)_{t\in{\bf Z}}$ where $\rho_t$ is a probability measure on $M$, such that 
$$	\rho_t=\lim_{n\to\infty}f^*_t\ldots f^*_{t-n}m\eqno{(1)}      $$
and the probability measure $m$ is absolutely continuous with respect to the volume element on $M$.  In particular $\rho_t=f^*_t\rho_{t-1}$.
\medskip
	Note that the above situation arises for random dynamical systems\footnote{*}{Such random dynamical systems have been much studied, see Kifer [55], Arnold [2], and for the SRB states Bahnm\"uller and Liu [3].} where a shift-invariant measure ${\bf P}(d\alpha)$ is given on the $\alpha=(f_t)_{t\in{\bf Z}}$, and statements are made for ${\bf P}$-almost every $\alpha$.  (The case of a perturbation proportional to $e^{i\omega t}$ that we shall consider later can be handled in this setup).  
\medskip
	For applications to physics, however, it is desirable to be able to consider a nonprobabilistic situation.  Secifically, we proceed with a brief discussion of the case where all $f_t$, $t\in{\bf Z}$, are sufficiently close to an Anosov diffeomorphism\footnote{*}{More generally one could consider an Axiom A diffeomorphism $f$ near an attractor, see [86], Section 4} $f$ of $M$.  Structural stability extends to this situation and gives a (unique) family $(j_t)_{t\in{\bf Z}}$ of homeomorphisms of $M$ close to the identity such that we have a commutative diagram
\def\fl#1{\uparrow\vbox to 5mm{}\rlap{$\scriptstyle #1$}}
$$\matrix{
\cdots&\rightarrow&M&\buildrel f_t\over{\rightarrow}&M
&\buildrel f_{t+1}\over{\rightarrow}&M&\rightarrow
&\cdots\cr&&\fl{j_{t-1}}&&\fl{j_t}&&\fl{j_{t+1}}&&\cr
\cdots&\rightarrow&M&\buildrel f\over{\rightarrow}&M
&\buildrel f\over{\rightarrow}&M&\rightarrow&\cdots\cr
}$$
	Using hyperbolicity, the fact that the $j_t$ are close to the identity, and $Tf_t$ close to $Tf$, one can define stable and unstable subbundles $V_t^s$, $V_t^u$ of $TM$ with the usual properties.  The corresponding (local) stable and unstable manifolds ${\cal V}_t^s(x)$, ${\cal V}_t^u(x)$ are smooth because we are in a uniformly hyperbolic situation.  One checks also readily that $j_t^{-1}{\cal V}_t^{s,u}(j_tx)$ coincides (in a sufficiently small neighborhood of $x$) with the stable or unstable manifold ${\cal V}^{s,u}(x)$ for $f$.  Note that although $j_t$ is not smooth in general, it maps the smooth manifolds ${\cal V}^{s,u}(x)$ to the smooth manifolds ${\cal V}_t^{s,u}(x)$.  Let us now define the time dependent SRB state $(\rho_t)$ by (1).  One can then show that the conditional measures of $\rho_t$ on the unstable manifolds ${\cal V}_t^u(x)$ are absolutely continuous with respect to the Riemann volume element of ${\cal V}_t^u(x)$ and satisfy a uniform continuity condition in $t$.  Conversely, if $(\rho_t)_{t\in{\bf Z}}$ satisfies $f_t^*\rho_{t-1}=\rho_t$ and the above absolute continuity and uniformity, then $(\rho_t)$ is a time dependent SRB state.  Such states can also be studied by transfer operator techniques (see [86]).
\medskip
	Although time dependent SRB states are conceptually important for nonequilibrium statistical mechanics, we shall mostly use those which correspond to an infinitesimal time dependent perturbation of a time independent dynamical system.  Also, we shall mostly use a continuous rather than discrete time, and work formally rather than rigorously.  
\medskip
	{\bf 3. A general linear response formula.}
\medskip
	Consider time dependent dynamics defined by
$$	{dx\over dt}=F_t(x)\eqno{(2)}      $$
on the compact manifold $M$.  Integrating (2) with initial condition $x$ at time $s$ gives at time $t$ a point $x(t,s)=f(t,s)x$.  For this time evolution, a time dependent SRB state $(\rho_t)_{t\in{\bf R}}$ is defined by 
$$	\rho_t=\lim_{s\to-\infty}f(t,s)^*m\eqno{(3)}      $$
where $m$ is absolutely continuous with respect to the volume element on $M$, and we assume that the limit (3) exists.
\medskip
	Let now $(\rho_t+\delta\rho_t)_{t\in{\bf R}}$ be the time dependent SRB state corresponding to the time evolution
$$	{dx\over dt}=F_t(x)+\delta F_t(x)      $$
we have thus
$$	\rho_t(\Phi)+\delta\rho_t(\Phi)=\lim_{s\to-\infty}
	m(\Phi\circ(f(t,s)+\delta f(t,s)))      $$
Proceeding formally this gives
$$	\delta\rho_t(\Phi)=\lim_{s\to-\infty}\int m(dx)
	(\nabla_{x(t,s)}\Phi)\cdot\delta x(t,s)      $$
and we may assume $s<t$.  Using the tangent map $T_xf$ to $f$ at $x$ we may write
$$	\delta x(t,s)
	=\int_s^t d\tau(T_{x(\tau,s)}f(t,\tau))\delta F_\tau(x(\tau,s))      $$
hence
$$	\delta\rho_t(\Phi)=\lim_{s\to-\infty}\int_s^td\tau\int m(dx)
((T_{x(\tau,s)}f(t,\tau))\delta F_\tau(x(\tau,s)))\cdot\nabla_{x(t,s)}\Phi $$
$$	=\lim_{s\to-\infty}\int_s^td\tau\int(f(\tau,s)^*m)(dy)
	((T_yf(t,\tau))\delta F_\tau(y))\cdot\nabla_{y(t,\tau)}\Phi      $$
$$	=\int_{-\infty}^td\tau\int\rho_\tau(dy)
	((T_yf(t,\tau))\delta F_\tau(y))\cdot\nabla_{y(t,\tau)}\Phi      $$
Finally
$$	\delta_t\rho(\Phi)=\int_{-\infty}^td\tau\int\rho_\tau(dy)
	\delta F_\tau(y)\cdot\nabla_y(\Phi\circ f(t,\tau))\eqno{(4)}      $$
$$	=\int_{-\infty}^td\tau\int\rho_t(dx)
((T_{x(\tau,t)}f(t,\tau))\delta F_\tau(x(\tau,t))\cdot\nabla_x\Phi\eqno{(5)}$$
This is the most general form of our linear response formula.
\medskip
	The convergence of the right-hand side of (4), (5) is assured if $\tau\to\delta F_\tau$ has compact support, but one would like to assume just boundedness, to allow time independent and time periodic perturbations.  As discussed above, one expects that (4), (5) can be proved under uniform hyperbolicity assumptions, but not in general.  As things stand, the case of Anosov diffeomorphisms and time independent perturbations has been treated rigorously, and one obtains the following result.
\medskip
	{\bf 4. Theorem} {\rm (Ruelle [86]).}
\medskip
	{\sl Let ${\cal A}$ be the space of $C^r$ Anosov diffeomorphisms of the compact connected manifold $M$, and let $\rho$ be the SRB state for $f\in{\cal A}$.  If $r\ge3$, the map $f\mapsto\rho|C^{r-1}(M)$ is $C^{r-2}:{\cal A}\to C^{r-1}(M)^*$.
\medskip
	The derivative of $f\mapsto\rho$ is given by
$$	\delta\rho(\Phi)
	=\sum_{n=0}^\infty\rho(X\cdot\nabla(\Phi\circ f^n))\eqno{(6)}      $$
where we have introduced the vector field $X=\delta f\circ f^{-1}$, and taken $\Phi\in C^{r-1}(M)$.}
\medskip
	Note that for time dependent dynamics $(f_t)_{t\in{\bf Z}}$ and time dependent SRB state $(\rho_t)$, (6) should be replaced by\footnote{*}{At the time of writing, there is no written proof of (7).}
$$	\delta_t\rho(\Phi)=\sum_{\tau=-\infty}^t\rho_\tau
(X_\tau\cdot\nabla(\Phi\circ f_t\circ\ldots\circ f_{\tau+1}))\eqno{(7)}     $$
which is the discrete version of (4).
\medskip
	The existing proof of Theorem 4 applies actually to Axiom A attractors, and even to general ``Axiom A basic sets'', provided the definition of SRB measures is generalized and (6) replaced by a more complicated formula (see [86]).
\medskip
	{\bf 5. Time dependent perturbation of a nonequilibrium steady state.}
\medskip
	We shall now assume that $F_t=F$ is time independent, and let $\rho_t=\rho$ be a time independent measure corresponding to the time evolution
$$	{dx\over dt}=F(x)\eqno{(8)}      $$
We may here write $f(t,s)=f^{t-s}$, $x(t,s)=x(t-s)=f^{t-s}x$.  The pertubation $\delta F_t$, noted now $\delta_t F$, will remain time dependent, so that the perturbed time evolution is 
$$	{dx\over dt}=F(x)+\delta_t F(x)\eqno{(9)}      $$
Let $(\rho+\delta_t \rho)$ be the time dependent SRB state for the evolution (9) which replaces the time independent SRB state $\rho$ for the evolution (8).  We are interested in the response $(\delta_t \rho)$ to the perturbation $(\delta_t F)$.  In view of (4), (5) we have (see [88])
$$	\delta_t\rho(\Phi)=\int_{-\infty}^td\tau\int\rho(dy)
	\delta_\tau F(y)\cdot\nabla_y(\Phi\circ f^{t-\tau})\eqno{(10)}      $$
$$	=\int_{-\infty}^td\tau\int\rho(dx)((T_{x(\tau-t)}f^{t-\tau})
	\delta_\tau F(x(\tau-t))\cdot\nabla_x\Phi\eqno{(11)}      $$
Assuming that $\Phi$ is in a suitable space ${\cal B}$ of functions on $M$, and $\delta F$ in a suitable space ${\cal X}$ of vector fields we may rewrite (10), (11) as 
$$	\delta_t\rho=\int d\tau\,\kappa_{t-\tau}\delta_\tau F
	=\int d\sigma\,\kappa_\sigma\delta_{t-\sigma}F      $$
where the linear operator $\kappa_\sigma$ maps ${\cal X}$ to the dual ${\cal B}^*$ of ${\cal B}$, and
$$	(\kappa_\sigma X)\Phi=0\qquad{\rm for}\qquad\sigma<0      $$
$$	(\kappa_\sigma X)\Phi=\int\rho(dx)((T_{x(-\sigma)}f^\sigma)
X(f^{-\sigma}x))\cdot\nabla_x\Phi\qquad{\rm for}\qquad\sigma\ge0\eqno{(12)} $$
\medskip
	{\bf 6. Properties of the response function.}
\medskip
	The (operator-valued) function $\sigma\mapsto\kappa_\sigma$ defined by (12) is called {\it response function}.  The fact that $\kappa_\sigma=0$ for $\sigma<0$ is called {\it causality} and is interpreted to mean that the cause $\delta_\tau F$ must precede the effect $\delta_t\rho$.
\medskip
	To study the behavior of $\kappa_\sigma$ for large $\sigma$ we shall assume a weak form of hyperbolicity, namely that all Lyapunov exponents\footnote{*}{See Section 2.5.} with respect to $\rho$ are $\ne0$, except one corresponding to the direction $F$ of the flow.  We may thus write for any vector field $X$ on $M$
$$	X=X^s+\phi F+X^u\eqno{(13)}      $$
where $X^s$ is in the stable (contracting), $X^u$ in the unstable (expanding) direction and $\phi$ is a scalar function.
\medskip
	{\bf Lemma.} {\rm (Ruelle [88]).} 
\medskip
	{\sl For $\sigma\ge0$ we have 
$$	(\kappa_\sigma X)\Phi=\int\rho(dy)
	[((T_yf^{\sigma})X^s(y))\cdot\nabla_{y(\sigma)}\Phi
	-(F\cdot\nabla\phi+{\rm div}^uX^u)(y)\Phi(y(\sigma))]\eqno{(14)}    $$
where the unstable divergence ${\rm div}^u$ will be defined below.}
\medskip
	We first obtain from (12), (13), 
$$	(\kappa_\sigma X)\Phi=\int\rho(dy)
	[((T_yf^{\sigma})X^s(y))\cdot\nabla_{y(\sigma)}\Phi
	+(\phi F+X^u)(y)\cdot\nabla_y(\Phi\circ f^\sigma)]      $$
To transform the $\phi$-term we use
$$	\phi F\cdot\nabla\Psi=\phi{d\over dt}\Psi\circ f^t|_{t=0}
	={d\over dt}((\phi\Psi)\circ f^t)|_{t=0}
	-({d\over dt}\phi\circ f^t|_{t=0})\Psi      $$
$$	={d\over dt}((\phi\Psi)\circ f^t)|_{t=0}-(F\cdot\nabla\phi)\Psi      $$
so that with $\Psi=\Phi\circ f^\sigma$ we get 
$$	\int\rho(dy)(\phi F)(y)\cdot\nabla_y(\Phi\circ f^\sigma)
	=-\int\rho(dy)(F\cdot\nabla\phi)(y)\Phi(y(\sigma))       $$
\medskip
	To transform the $X^u$-term we shall use the fact that $\rho$ is an SRB measure.  We may write for $\rho$ a disintegration of the form
$$	\rho(dy)=\int\nu(d\lambda)\rho_{(\lambda)}(dy)      $$
where $\rho_{(\lambda)}(dy)$ is a conditional measure on a local unstable manifold.  The SRB condition is that $\rho_{(\lambda)}(dy)$ is absolutely continuous with respect to the Riemann volume element $d_{(\lambda)}^u(dy)$ on the (local) unstable manifold labelled by $\lambda$.  Writing $\rho_{(\lambda)}(dy)=m_{(\lambda)}(y)d_{(\lambda)}^u(dy)$ one sees that the density $m_{(\lambda)}(\cdot)$ is unique up to a multiplicative constant, in fact 
$$	{m(x)\over m(y)}=\prod_{n=1}^\infty{J^u(f^{-n}x)\over J^u(f^{-n}y)}  $$
where $J^u$ denotes the unstable Jacobian\footnote{*}{See Section 2.4.} for $f$.  Let us place ourselves in the unstable manifold indexed by $\lambda$.  Since $\rho_{(\lambda)}(dy)$ is a volume element, its Lie derivative with respect to a vector field $Y$ satisfies 
$$	L_Y\rho_{(\lambda)}=({\rm div}^uY)\rho_{(\lambda)}      $$
where ${\rm div}^uY$ is the divergence of $Y$ with respect to the volume element $\rho_{(\lambda)}(dy)$.  Note that ${\rm div}^u$ does not depend on the choice of arbitrary multiplicative constant in the definition of $\rho_{(\lambda)}(dy)$.  We call ${\rm div}^u$ the {\it unstable divergence} and obtain
$$	\int\rho_{(\lambda)}(dy)(Y\cdot\nabla\Psi)(y)
	=-\int\rho_{(\lambda)}(dy)({\rm div}^uY)(y)\Psi(y)
	+\hbox{ boundary term}      $$
Since $X^u$ is tangent to the unstable manifolds we have 
$$	\int\rho(dy)X^u(y)\cdot\nabla_y\Psi
	=\int\nu(d\lambda)\int\rho_{(\lambda)}(dy)X^u(y)\cdot\nabla_y\Psi    $$
$$	=-\int\nu(d\lambda)\int\rho_{(\lambda)}(dy)({\rm div}^uX^u)(y)\Psi(y)
	=-\int\rho(dy)({\rm div}^uX^u)(y)\Psi(y)      $$
where the boundary terms disappear because of cancellation between the local pieces of a global unstable manifold.  Taking $\Psi=\Phi\circ f^\sigma$ we obtain thus
$$	\int\rho(dy)X^u(y)\cdot\nabla_y(\Phi\circ f^\sigma)
	=-\int\rho(dy)({\rm div}^uX^u)(y)\Phi(y(\sigma))      $$
which completes the proof of (14).\qed
\medskip
	{\bf 7. Physical discussion: the question of integrability of the response function.}
\medskip
	We interrupt here for a moment our technical discussion, and we pause to consider the relation of our analysis of linear response with more traditional approaches.  (For the latter see for instance de Groot and Mazur [20]).
\medskip
	The usual approaches usually suppose closeness to equilibrium (as discussed in Chapter 5 below) and assume that the response of a system to small perturbations is empirically given.  Physical principles such as causality or microscopic reversibility are then seen to have consequences for the response, and these consequences are investigated.
\medskip
	In the analysis presented here, by contrast, the response of a system to perturbations is deduced from the general principle that nonequilibrium steady states are given by SRB measures.  This new approach, being much more explicit, applies directly far from equilibrium, and causality is a consequence of the choice of SRB measures instead of appearing as an extra assumption.
\medskip
	In the traditional approaches it appears reasonable to assume that the empirically defined response function is well behaved, say integrable.  In the new approach this has to be discussed on the basis of expressions such as (12) or (14), and we must admit that we are on somewhat shaky ground.  {\it Formally,} we can argue as follows.  Since $X^s$ is in the contracting direction, the term
$$	\int\rho(dy)((T_yf^{\sigma})X^s)\cdot\nabla_{y(\sigma)}\Phi      $$
should decrease exponentially with $\sigma$.  The term
$$	\int\rho(dy)(F\cdot\nabla\phi+{\rm div}^uX^u)(y)\Phi(f^\sigma y)     $$
is a correlation function (note that $\int\rho(dy)(F\cdot\nabla\phi+{\rm div}^uX^u)(y)=0$), and we may hope that it decreases exponentially when $\sigma\to\infty$.  Therefore $(\kappa_\sigma X)\Phi$ is an exponentially decreasing function of $\sigma$, and is in particular integrable.
\medskip
	In fact difficulties arise because the terms $X^s$, $\phi F$, $X^u$ of the decomposition (13) of $X$ need not be continuous functions on $M$ unless we assume uniform hyperbolicity, and because even for Anosov flows the exponential decrease of correlations has been proved only under certain conditions (see Dolgopyat [25], [26]).  That these difficulties are not visible in the traditional approaches is because one is satisfied there with an empirical definition of the response function.  As soon as one tries to obtain a dynamical understanding of the response, difficulties necessarily arise.  A resolution of these difficulties will require a better understanding of smooth dynamics, and a better understanding of the infinite system limit (thermodynamic limit).
\medskip
	While we do not underestimate the difficulties just discussed, we shall want to push ahead, assuming that the response function may somehow be considered as integrable, as is done without trepidation in the more traditional approaches.  The physical interest of the results obtained may then justify further concern for their mathematical justification.
\medskip
	{\bf 8. Dispersion relations.}
\medskip
	The Fourier transform of the response function is called the {\it susceptibility}:
$$	\hat\kappa_\omega=\int\kappa_\sigma e^{i\omega\sigma}d\sigma      $$
Remember that $\kappa_\omega$, $\hat\kappa_\omega$ are linear operators ${\cal X}\to{\cal B}^*$.  Choosing $X\in{\cal X}$ and $\Phi\in{\cal B}$, we note that $(\hat\kappa_\omega X)\Phi$ is the response of a system in the nonequilibrium steady state $\rho$ to a periodic perturbation.
\medskip
	Taking $\sigma\mapsto(\kappa_\sigma X)\Phi$ to be integrable as discussed in Section 7, we see that $\omega\mapsto(\hat\kappa_\omega X)\Phi$ is continuous and tends to 0 at infinity (Riemann-Lebesgue lemma).  Furthermore, the fact that $\kappa_\sigma=0$ for $\sigma<0$ implies that $\omega\mapsto(\hat\kappa_\omega X)\Phi$ extends to an analytic function in the upper half complex plane.  The analyticity of $\omega\mapsto(\hat\kappa_\omega X)\Phi/(\omega-u)$ and standard manipulations (contour integration) yield then
$$	(\hat\kappa_u X)\Phi={1\over\pi i}\,\hbox{p.v.}
\int_{-\infty}^\infty{(\hat\kappa_\omega X)\Phi\over\omega-u}\,d\omega      $$
where p.v. denotes principal value.  More formally
$$	\hat\kappa_u={1\over\pi i}\,\hbox{p.v.}
\int_{-\infty}^\infty{\hat\kappa_\omega\over\omega-u}\,d\omega\eqno{(15)}    $$
Splitting $\hat\kappa$ into real and imaginary parts, one can also write that one is the Hilbert transform of the other.  The functional identity (15) is the {\it Kramers-Kronig dispersion relation}.  Note that no condition of closeness to equilibrium was necessary to derive (15).
\medskip
	{\bf 9. Correlation functions, spectral density and resonances.}
\medskip
	Let $\rho$ be any probability measure on the compact space $M$, invariant under the time evolution $(f^t)_{t\in{\bf Z}}$, or $(f^t)_{t\in{\bf R}}$.  There is a group homomorphism $U$ of ${\bf Z}$ or ${\bf R}$ into the unitary operators on $L^2(M,\rho)$, defined by 
$$	U(t)\Psi=\Psi\circ f^t      $$
There is also a homomorphism $\alpha$ of the algebra ${\cal C}(M,{\bf C})$ of complex continuous functions on $M$ into the algebra ${\cal B}(L^2(M,\rho))$ of bounded operators on $L^2(M,\rho)$, defined by 
$$	(\alpha(A)\Psi)(x)=A(x)\Psi(x)      $$
The homomorphism $\alpha$ satisfies $\alpha(A^*)=(\alpha A)^*$ (adjoint operator) and $\alpha(1)={\bf 1}$ (unit operator).  Furthermore
$$	\alpha(A\circ f^t)=U(t)\alpha(A)U(-t)      $$
and if $\Omega$ is the function $1$ considered as element of $L^2(M,\rho)$ we have\footnote{*}{The construction just indicated is a special case of the GNS construction (Gel'fand-Na\v\i mark-Segal).}
$$	U(t)\Omega=\Omega      $$
$$	(\Omega,\alpha(A)\Omega)=\rho(A)      $$
\par
	If $A$, $B\in{\cal C}(M,{\bf C})$ we define the (time) correlation fuction
$$	\rho_{AB}(t-s)=\rho((A\circ f^s)(B\circ f^t))-\rho(A)\rho(B)      $$
$$	=(\alpha(A^*)\Omega,U(t-s)\alpha(B)\Omega)
	-(\Omega,\alpha(A)\Omega)(\Omega,\alpha(B)\Omega)      $$
Introduce the spectral representation 
$$	U(t)=\int e^{i\omega\tau}{\bf P}(d\omega)      $$
where ${\bf P}$ is the spectral measure on the circle ${\bf R}({\rm mod}\enspace2\pi)$ in the discrete time case or on ${\bf R}$ in the continuous time case.  Note that ${\bf P}(\{0\})$ is the orthogonal projection on the space ${\cal H}_0$ of $U(t)$-invariant vectors (for all $\tau\in{\bf N}$ or ${\bf R}$), and that $\Omega\in{\cal H}_0$.  Ergodicity of the measure $\rho$ with respect to $(f^t)$ is equivalent to $\dim{\cal H}_0=1$ ({\it i.e.}, there are no nontrivial invariant functions).  Assuming ergodicity we have thus
$$	{\bf P}(\{0\})=\hbox{orthogonal projection $P_\Omega$ on $\Omega$}  $$
Writing $\Psi_1=\alpha(A^*)\Omega$, $\Psi_2=\alpha(B)\Omega$, we obtain
$$	\rho_{AB}(\tau)=\rho(A(B\circ f^\tau))-\rho(A)\rho(B)
	=(\Psi_1,U(\tau)\Psi_2)-(\Psi_1,P_\Omega\Psi_2)      $$
$$   =\int_{\{\omega\ne0\}}e^{i\omega\tau}(\Psi_1,{\bf P}(d\omega)\Psi_2)   $$
In particular, if $A=A^*$ and $\Psi=\alpha(A)\Omega$ we find 
$$   \rho_{AA}(\tau)=\int e^{i\omega\tau}\,\check\rho_{AA}(\omega)\,d\omega  $$
where 
$$	(\Psi,{\bf P}(d\omega)\Psi)
	=(\check\rho_{AA}(\omega)+|(\Psi,\Omega)|^2\delta(\omega))d\omega    $$
and $\check\rho_{AA}(\omega)\ge0$.  The time autocorrelation function $\rho_{AA}$ is thus the Fourier transform of the positive density $\check\rho_{AA}(\omega)$, called {\it spectral density}, or {\it power spectrum} (of the ``signal'' $t\mapsto A\circ f^t$).
\medskip
	Let us now specialize to the situation where $\rho$ is an SRB measure for a smooth dynamical system $(f^t)$ on a compact manifold.  Under various hyperbolicity assumptions (in particular uniform hyperbolicity) and under suitable smoothess assumptions on $A$, $B$, one can show that $\rho_{AB}$ is the Fourier transform of a continuous function $\omega\to\check\rho_{AB}(\omega)$ which extends meromorphically to $|{\rm Im}\omega|<c$ for some constant $c>0$.  The poles of $\rho_{AB}$ are called {\it resonances}, their position is independent of the choice of $A$, $B$ (although some residue might accidentally vanish).  For a discussion of resonances see in particular Ruelle [83], [84], Parry and Pollicott [70], Baladi [4], Dolgopyat and Pollicott [27].
\medskip
	{\bf 10. Fluctuation-Dissipation.}
\medskip
	In nonequilibrium statistical mechanics close to equilibrium there is an important result known as fluctuation-dissipation theorem (see Chapter 5).  We discuss here briefly and formally what remains of this result far from equilibrium on the basis of the following lemma.
\medskip
	{\bf Lemma.}  
\medskip
	{\sl The susceptibility may be written as 
$$	\hat\kappa_\omega=\hat\kappa^s_\omega+\hat\kappa^u_\omega      $$
where
$$	(\hat\kappa^s_\omega X)\Phi=\rho[(\int_0^\infty e^{i\omega\sigma}
	\,d\sigma((Tf^\sigma)X^s)\circ f^{-\sigma})\cdot\nabla\Phi]      $$
$$	(\hat\kappa^u_\omega X)\Phi=-\int_0^\infty e^{i\omega\sigma}\,d\sigma
	\rho[((F\cdot\nabla\phi+{\rm div}^uX^u)\circ f^{-\sigma})\Phi]      $$}
This follows directly from the lemma of Section 6.\qed
\medskip
	The functions $\omega\mapsto(\hat\kappa^s_\omega X)\Phi$, $(\hat\kappa^u_\omega X)\Phi$ extend analytically to the upper half-plane by causality.  Formally, the function $\sigma\mapsto\rho[((F\cdot\nabla\phi+{\rm div}^uX^u)\circ f^{-\sigma})\Phi]$ is a correlation function (note that $\rho(F\cdot\nabla\phi+{\rm div}^uX^u)=0$); therefore we may hope that its Fourier transform extends meromorphically to a strip $|{\rm Im}\omega|<c$ as discussed in Section 9.  This implies that $\omega\mapsto(\hat\kappa^u_\omega X)\Phi$ extends meromorphically to ${\rm Im}\omega>-c$, its poles for $0>{\rm Im}\omega>-c$ being the same resonances as those of the spectral density.
\medskip
	Formally, we also obtain that $(\hat\kappa^s_\omega X)\Phi$ is holomorphic for ${\rm Im}\omega>\lambda$ where $\lambda$ is the largest negative characteristic exponent of $(Tf^\sigma)$ with respect to $\rho$.  We can also define ${\cal T}^\sigma$ on the vector fields in the stable direction by
$$	(Tf^\sigma Y)(x)=(T_{f^{-\sigma}x}f^\sigma)Y(f^{-\sigma}x)      $$
Then, on a suitable Banach space, $({\cal T}^\sigma)_{\sigma\ge0}$ is a contraction semigroup and, if $-H$ is its infinitesimal generator we have by th Hille-Yosida theorem
$$	\int_0^\infty e^{i\omega\sigma}d\sigma((Tf^\sigma)Y)\circ f^{-\sigma}
	=(H-i\omega)^{-1}Y      $$
so that $(\hat\kappa^s_\omega X)\Phi=\rho[((H-i\omega)^{-1}X^s)\cdot\nabla\Phi]$.  The singularities of $\hat\kappa^s_\omega$ are thus related to the spectrum of $H$.
\medskip
	In conclusion the susceptibility has {\it stable} and {\it unstable} poles, which can in principle be distinguished experimentally because the unstable poles are the same as those of a spectral density.  Close to equilibrium the difference between these two kinds of poles disappears as we shall see in the next chapter.
\vfill\eject

\centerline{Chapter 5. NONEQUILIBRIUM NEAR EQUILIBRIUM.}
\bigskip\bigskip
	{\bf 1. Terminology and assumptions.}
\medskip
	An isolated macroscopic system which is not subjected to external forces normally evolves towards a macroscopic state of rest\footnote{*}{We assume for simplicity that the total momentum and angular momentum vanish.} corresponding to certain values of thermodynamic variables like the temperature.  At the macroscopic level, this state of thermodynamic equilibrium is described by the {\it microcanonical ensemble} of equilibrium statistical mechanics (see Section 1.9).  Typically, we have a phase space ${\bf R}^{2N}$ with a Hamiltonian function $({\bf p},{\bf q})\mapsto H({\bf p},{\bf q})$ where ${\bf p}=(p_1,\ldots,p_N)$, ${\bf q}=(q_1,\ldots,q_N)$, and the so-called microcanonical ensemble corresponding to a certain energy $K$ is the probability measure 
$$	\rho(d{\bf p}d{\bf q})
	={1\over\Omega}\delta(H({\bf p},{\bf q})-K)\eqno(1)      $$
where $\Omega$ is a normalizing factor and it is assumed (typically) that $M=\{({\bf p},{\bf q})\in{\bf R}^{2N}:H({\bf p},{\bf q})=K\}$ is compact and that ${\rm grad}H$ does not vanish on $M$, so that $M$ is a smooth compact manifold.
\medskip
	Note that at the microscopic level we do not have rest or equilibrium, but a nontrivial time evolution
$$	{d\over dt}\pmatrix{{\bf p}\cr{\bf q}\cr}
	=\pmatrix{-\partial_{\bf q}H({\bf p},{\bf q})\cr
	\partial_{\bf p}H({\bf p},{\bf q})\cr}\eqno{(2)}     $$
We shall nevertheless use the traditional terminology and call {\it equilibrium} the situation where we have a smooth time evolution $(f^t)$ (defined by (2)) on the compact manifold $M$, and a probability measure $\rho$ (defined by (1)), invariant under $(f^t)$, and with smooth nonvanishing density with respect to the Riemann volume element $dx$ on $M$.  It is desirable to assume that $\rho$ is ergodic.  In physical applications, $\rho$ may have one large ergodic component, with the measure of the remainder tending to zero in the {\it thermodynamic limit} ({\it i.e.}, for large systems); it would not be difficult to adapt our discussion to this situation.  In the present chapter we shall disregard the symplectic structure implied by the Hamiltonian time evolution $(f^t)$ (symplectic structure will be important for the Dettmann-Morriss pairing theorem, see Chapter 6).
\medskip
	Notice that the equilibrium state $\rho$ is an SRB measure.  In fact, for the purposes of this chapter, an {\it equilibrium state} $\rho$ may be defined as an SRB measure with smooth density ($\ne0$) with respect to $dx$, the time being allowed to be discrete ($t\in{\bf Z}$) or continuous ($t\in{\bf R}$).  We shall see in Section 3 that the entropy production vanishes for an equilibrium state, which is therefore described as {\it nondissipative}.  (An SRB state with entropy production $>0$ is {\it dissipative}).  In Section 7 we shall assume the reversibility of the dynamics ({\it i.e.}, there is a diffeomorphism $i:M\to M$ such that $i^2={\rm identity}$, $if^t=f^{-t}i$), and of the equilibrium state $\rho$ ({\it i.e.}, $i^*\rho=\rho$)\footnote{**}{In Chapter 3 we assumed reversibility of the dynamics, but the ``anti-SRB'' state $i^*\rho$ was different from $\rho$.  The condition $i^*\rho=\rho$ can be imposed only at equilibrium.}.  Actually the latter condition follows from the assumed ergodicity: $i^*\rho$ is absolutely continuous with respect to $dx$, hence with respect to the ergodic measure $\rho$, and since $i^*\rho$ is $(f^t)$-invariant, $i^*\rho=\rho$.
\medskip
	To summarize: {\sl We are given a smooth dynamical system $(f^t)_{t\in{\bf Z}}$ or $(f^t)_{t\in{\bf R}}$ on the compact manifold $M$, with an ergodic probability measure $\rho$ called {\rm equilibrium state}.  We assume that $\rho(dx)$ has smooth nonvanishing density with respect to $dx$.  In Section 7 we also assume reversibility, {\it i.e.}, there is a diffeomorphism $i$ of $M$ such that $i\circ i={\rm identity}$, $i\circ f^t=f^{-t}\circ i$; in particular this implies $i^*\rho=\rho$.}
\medskip
	These assumptions will allow us to recover the results of nonequilibrium statistical mechanics close to equilibrium.  We shall be satisfied in this chapter with {\it formal} derivations (for the rigorous treatment of uniformly hyperbolic dynamical systems, see Chapter 5).  The rest of this chapter is based in good part on Gallavotti and Ruelle [46].
\medskip
	{\bf 2. Linear response close to equilibrium.}
\medskip
	Given an SRB state $\rho$ for $(f^t)$, we have introduced in Chapter 4 the response function $\sigma\mapsto\kappa_\sigma$ such that $\kappa_\sigma=0$ for $\sigma<0$ and
$$	(\kappa_\sigma X)\Phi
=\int\rho(dx)((T_{x(-\sigma)}f^\sigma)X(f^{-\sigma}x))\cdot\nabla_x\Phi      $$
$$	=\int\rho(dy)X(y)\cdot\nabla_y(\Phi\circ f^\sigma)
	\qquad{\rm for}\qquad\sigma\ge0\eqno{(3)}      $$
With this definition we have shown that the linear change $\delta_t\rho$ of the SRB state $\rho$ in response to a change $\delta_t F$ in the force $F$ acting on the system is given by
$$	\delta_t\rho=\int d\sigma\kappa_\sigma\delta_{t-\sigma}F\eqno{(4)}   $$
in the continuous time case ($t\in{\bf R}$).  In the discrete time case ($t\in{\bf Z}$) we have with the same definition of $\kappa_\sigma$
$$	\delta_t\rho
=\sum_\sigma\kappa_\sigma(\delta_{t-\sigma}f\circ f^{-1})\eqno{(5)}      $$
where $\delta_tf$ is the infinitesimal change of $f$ at time $t$ (and $\delta_tf\circ f^{-1}$ is thus an infinitesimal vector field on $M$).
\medskip
	We shall denote by ${\rm div}V$ the divergence of a vector field $V$ on $M$ with respect to the volume element $\rho(dx)$.  By definition, the Lie derivative of this volume element satisfies
$$	L_V\rho=({\rm div}V)\rho      $$
We may thus rewrite (3) as
$$	(\kappa_\sigma X)\Phi=\int\rho(dy)X(y)\cdot\nabla_y(\Phi\circ f^\sigma)
	=-\int\rho(dy){\rm div}X(y).\Phi(f^\sigma y)\eqno{(6)}      $$
for $\sigma\ge0$.
\medskip
	For the susceptibility we obtain in the continuous time case ($\omega\in{\bf R}$):
$$	(\hat\kappa_\omega X)\Phi=\int_0^\infty e^{i\omega\sigma}d\sigma
	\int\rho(dy)X(y)\cdot\nabla_y(\Phi\circ f^\sigma)      $$
$$	=-\int_0^\infty e^{i\omega\sigma}d\sigma\int\rho(dy)
	{\rm div}X(y).\Phi(f^\sigma y)      $$
and in the discrete time case ($\omega\in{\bf R}(\hbox{mod }2\pi)$)
$$	(\hat\kappa_\omega X)\Phi=\sum_{\sigma=0}^\infty e^{i\omega\sigma}
	\int\rho(dy)X(y)\cdot\nabla_y(\Phi\circ f^\sigma)      $$
$$	=-\sum_{\sigma=0}^\infty e^{i\omega\sigma}\int\rho(dy)
	{\rm div}X(y).\Phi(f^\sigma y)      $$
In particular for time independent $\delta F$ or $\delta f$ we have respectively
$$	(\delta\rho)\Phi=(\hat\kappa_0\delta F)\Phi
	=-\int_0^\infty d\sigma\int\rho(dy){\rm div}\delta F(y)
	.\Phi(f^\sigma y)      $$
or
$$	(\delta\rho)\Phi=(\hat\kappa_0(\delta f\circ f^{-1}))\Phi
=-\sum_{\sigma=0}^\infty\int\rho(dy)({\rm div}(\delta f\circ f^{-1}))(y)
	.\Phi(f^\sigma y)      $$
\medskip
	{\bf 3. Entropy production.}
\medskip
	Suppose that $\rho$ is an equilibrium state corresponding to the vector field $F$ (or the diffeomorphism $f$).  For a small time dependent change $\delta_t F$ of $F$ (or $\delta_t f$ of $f$) we shall be able to determine the entropy production {\it to second order}.
\medskip
	Remember that in a time independent situation we can (as noted in Section 1.9) use any smooth volume element to compute the divergence (or the Jacobian) occuring in the definition of the entropy production.  To estimate a time dependent entropy production we have however to make a definite choice of volume element. We take here $\rho(dx)$, which is the natural choice.  In the continuous time case we have then
$$	{\rm div}F=0      $$
[Indeed, by definition, $\rho(dx){\rm div}F(x)=L_F\rho(dx)=0$ because $\rho$ is invariant under the time evolution defined by $F$].  In the discrete time case we have
$$	J=1      $$
[because $f$ preserves $\rho(dx)$].  We expand the SRB state corresponding to $F+\delta_t F$ (or $f+\delta_t f$) to second order as $\rho+\delta_t\rho+\delta^{(2)}_t\rho$, and compute the corresponding entropy production.  The term corresponding to $\delta^{(2)}\rho$ does not contribute to the result, which is as follows.
\medskip
	{\bf Proposition} (time dependent entropy production).
\medskip
	{\sl In the continuous time case, writing $X_t=\delta_t F$, the entropy production is to second order
$$	e_t=\int_0^\infty d\sigma\int\rho(dy)
	{\rm div}X_{t-\sigma}(y).{\rm div}X_t(f^\sigma y)\eqno{(7)}      $$
In the discrete time case, writing $X_t=\delta_t f\circ f^{-1}$, the entropy production is to second order
$$	e_t={1\over2}\int\rho(dy)({\rm div}X_t(y))^2
	+\sum_{\sigma=1}^\infty\int\rho(dy)
	{\rm div}X_{t-\sigma}(y).{\rm div}X_t(f^\sigma y)\eqno{(8)}      $$}
\indent
	In the continuous time case we have to second order
$$	e_t=-\int(\rho(dx)+\delta_t\rho(dx)+\delta^{(2)}_t\rho(dx))
	({\rm div}F(x)+{\rm div}\delta_t F(x))      $$
Since ${\rm div}F=0$, the $\delta^{(2)}\rho$-term does not contribute, and since the $\rho$-integral of a divergence vanishes we find
$$	e_t=-\int\delta_t\rho(dx){\rm div}\delta_t F(x)
	=-(\delta_t\rho)({\rm div}X_t)      $$
Using the linear response (=first order) formulae (4), (6) gives then (7).
\medskip
	In the discrete time case, for the entropy production between time $t-1$ and time $t$, we have
$$	e_t=-\int(\rho(dy)+\delta_{t-1}\rho(dy)+\delta^{(2)}_{t-1}\rho(dy))
	\log J_{f+\delta_t f}(y)      $$
where 
$$	J_{f+\delta_t f}(y)=J_{{\rm identity}+\delta_t f\circ f^{-1}}(fy)
	=1+\hbox{higher order}      $$
so that the  $\delta^{(2)}\rho$-term vanishes.  Let us now suppose that the vector field $X_t=\delta_t f\circ f^{-1}$ has support covered by a local chart on which $\rho$ is proportional to Lebesgue measure: $\rho(dx)=c\,dx$.  Using the coordinates of this chart, and denoting by $A$ the matrix $(\partial_iX_j)$, with $X=X_t$, we have to second order
$$      \int\rho(dy)\log J_{f+\delta_t f}(y)
	=\int\rho(dx)\log J_{{\rm identity}+X_t}(x)      $$
$$	=c\int dx\log\det({\bf 1}+A)=c\int dx\,{\rm tr}\log({\bf 1}+A)      $$
$$	=c\int dx\,{\rm tr}(A-{1\over2}A^2)=c\int dx\,(\sum_i\partial_iX_i(x)
	-{1\over2}\sum_{ij}(\partial_iX_j(x))(\partial_jX_i(x)))      $$
Using integration by part, this is
$$	=-{c\over2}\int dx(\sum_i\partial_iX_i(x))^2
	=-{1\over2}\int\rho(dy)({\rm div}X(y))^2      $$
For general $X$ we may write $X=\sum X_\alpha$ and assume that every $|X_\alpha|+|X_\beta|$ has support as above; from this one obtains that the formula
$$	\int\rho(dy)\log J_{f+\delta_t f}(y)
	=-{1\over2}\int\rho(dy)({\rm div}X_t(y))^2      $$
holds generally (to second order).
\medskip
	Also to second order we have
$$	\int\delta_{t-1}\rho(dy)\log J_{f+\delta_t f}(y)
	=\int\delta_{t-1}\rho(dy)\log J_{{\rm identity}+X_t}(fy)      $$
and therefore
$$	e_t={1\over2}\int\rho(dy)({\rm div}X_t(y))^2 
	-\int\delta_{t-1}\rho(dy){\rm div}X_t(fy)     $$
Using the linear response (=first order) formulae (5), (6) gives finally
$$	e_t={1\over2}\int\rho(dy)({\rm div}X_t(y))^2
	+\sum_{\sigma=1}^\infty\int\rho(dy){\rm div}X_{t-\sigma}(y)
	.{\rm div}X_t(f^\sigma y)     $$
which proves (8).\qed
\medskip
	{\bf Corollary} (time independent entropy production).
\medskip
	{\sl If $X$ does not depend on $t$,
$$	e={1\over2}\int_{-\infty}^\infty d\sigma
	\int\rho(dx){\rm div}X(y).{\rm div}X(f^\sigma y)
	={1\over2}\int_{-\infty}^\infty d\sigma\,
	\rho_{{\rm div}X,{\rm div}X}(\sigma)      $$
in the continuous time case, and
$$	e={1\over2}\sum_{\sigma=-\infty}^\infty
	\int\rho(dx){\rm div}X(y).{\rm div}X(f^\sigma y)
	={1\over2}\sum_{\sigma=-\infty}^\infty
	\rho_{{\rm div}X,{\rm div}X}(\sigma)      $$
in the discrete time case.}
\medskip\noindent
[Remember that 
$\rho_{AA}(\sigma)=\int\rho(dy)A(y)A(f^\sigma y)-(\int\rho(dy)A(y))^2
,\int\rho(dy){\rm div}X(y)=0$].
\medskip
	{\bf Remark} (frequency dependent entropy production).
\medskip
	In the continuous time case, taking $X_t(y)=\sqrt{2}\sin{(\omega t-\theta)}X(y)$ and averaging over $t$ we obtain
$$	\langle e\rangle(\omega)=\int_0^\infty\cos{\omega\sigma}d\sigma
	\int\rho(dy){\rm div}X(y).{\rm div}X(f^\sigma y)      $$
$$	={1\over2}\int_{-\infty}^\infty e^{i\omega\sigma}d\sigma
	\rho_{{\rm div}X,{\rm div}X}(\sigma)      $$
$$	={1\over2}\hat\rho_{{\rm div}X,{\rm div}X}(\omega)
	=\pi\check\rho_{{\rm div}X,{\rm div}X}(\omega)\ge0      $$
[The right-hand side is proportional to the spectral density associated with the signal ${\rm div}X\circ f$ and is thus $\ge0$, see Section 4.9].
\medskip
	{\bf 4. Forces and observables close to equilibrium.}
\medskip
	In the last two sections we have studied the linear response and entropy production corresponding to a {\it force} $X$ ($=\delta F$ or $\delta f\circ f^{-1}$) pushing a system away from equilibrium.  A striking feature of the formulae that we have obtained is that $X$ occurs only in the form\footnote{*}{Remember that ${\rm div}X$ is the divergence of $X$ with respect to the volume element $\rho(dx)$ corresponding to the equilibrium state $\rho$.} ${\rm div}X$.  Nonequilibrium statistical mechanics close to equilibrium is very special because of this feature, which does not persist far from equilibrium. 
\medskip
	Let us call {\it observable} a function $\Phi:M\to{\bf R}$.  We further assume that $\Phi$ is smooth and that $\rho(\Phi)=0$.  Thus, if $X$ is any smooth vector field, ${\rm div}X$ is an observable, and every observable is of the form ${\rm div}X$.  [More precisely, {\it if $M$ is connected}, every function with a vanishing integral can be written as a divergence].  From a physical viewpoint, it is essential to be able to identify the observable ${\rm div}X$ correponding to a force $X$.  Remember that we typically deal with macroscopic systems, for which the mathematical definition of the divergence has no direct operational meaning.
\medskip
	To identify ${\rm div}X$ as an observable we shall use the Hilbert space
$$	{\cal H}=\{\Phi\in L^2(M,\rho):\rho(\Phi)=0\}      $$
The observables $\Phi$ belong to ${\cal H}$ and the scalar product $(\Psi,\Phi)$ is $\rho(\Psi\Phi)=\rho_{\Psi\Phi}(0)$ (value at 0 of the correlation function $\rho_{\Psi\Phi}(t)$.  We have in principle an operational definition of the scalar product
$$	(-{\rm div}X,\Phi)=-\int\rho(dx)\,{\rm div}X(x).\Phi(x)
	=\int\rho(dx)\,X\cdot\nabla\Phi(x)      $$
Indeed, by (4), (6) this is the response at time $0+$ of a time dependent force $X_t=\delta(t).X$ where $\delta(t)$ is the Dirac delta in the continuous time case (the discrete time case is similar).  The force $X_t$ is a kick at time 0, and $(-{\rm div}X,\Phi)$ is the {\it instantaneous} response of the system\footnote{**}{In the continuous time case, one could also switch on a constant force $X$ at time 0 and obtain $(-{\rm div}X,\Phi)$ as derivative of the expectation of $\Phi$ at $t=0+$.  In the discrete time case, $(-{\rm div}X,\Phi)$ is the jump in the expectation value of $\Phi$ when $f$ is replaced by $({\rm identity}+X)\circ f$.}.  One can also say that the kick $X_t$ replaces $\rho(dx)$ by $(1-{\rm div}X)\rho(dx)$ at time $0+$, where the new term $-{\rm div}X\rho(dx)$ may be interpreted as a {\it fluctuation}.  [We do not wish to pursue the discussion of fluctuations on the basis of Markovian and Gaussian assumptions as done for instance in [20]].
\medskip
	Notice that we can also obtain the scalar product $({\rm div}X,{\rm div}Y)$ by polarization of 
$$	({\rm div}X,{\rm div}X)=\rho(({\rm div}X)^2)
	={1\over\pi}\int\langle e\rangle(\omega)d\omega      $$
\indent
	The formulae written above permit in principle to identify ${\rm div}X$ as an observable.  From now on we accept without further discussion this identification, which is important in particular for the Onsager reciprocity relations and the fluctuation-dissipation theorem.
\medskip
	{\bf 5. Thermodynamic forces and conjugate fluxes.}
\medskip
	Let us assume that the dynamics $(f^t)$ of our system depends (smoothly) on parameters $E_\alpha$, as is natural in physical applications.  We also assume that when the parameters $E_\alpha$ take the value 0 equilibrium is achieved, corresponding to a vector field $F$ (continuous time case) or a diffeomorphism $f$ (discrete time case).  Close to equilibrium $F$ is replaced by $F+\delta F$, or $f$ by $f+\delta f$, and we take as earlier $X=\delta F$ or $X=\delta f\circ f^{-1}$.  We may thus write, to first order 
$$	X=\sum_\alpha V_\alpha E_\alpha      $$
({\it i.e.}, $\delta F=\sum_\alpha V_\alpha \delta E_\alpha$ or $\delta f\circ f^{-1}=\sum_\alpha V_\alpha \delta E_\alpha$, since $E_\alpha=\delta E_\alpha$).  We assume here that the $E_\alpha$ are time independent; the time dependent case is discussed in Section 6.  The {\it thermodynamic forces} which occur in the formalism of nonequilibrium thermodynamics (see [74], [20]) will now be identified with the $E_\alpha$.
\medskip
	In the continuous time case we define, following Gallavotti [37], [38], the {\it thermodynamic flux} conjugate to $E_\alpha$ as
$$	{\cal I}_\alpha=\rho_{F+\delta F}({\partial\over\partial E_\alpha}
	\sigma_{F+\delta F})      $$
where 
$$	\sigma_{F+\delta F}=-{\rm div}({F+\delta F})=-{\rm div}X      $$
is the local entropy production.  To first order in $X$ the flux is (using (4), (6))
$$	{\cal I}_\alpha=\delta\rho(-{\rm div}V_\alpha)
	=\int_0^\infty d\sigma\int\rho(dy)\,{\rm div}X(y)
.{\rm div}V_\alpha(f^\sigma y)=\sum_\beta L_{\alpha\beta}E_\beta\eqno{(9)}      $$
where 
$$	L_{\alpha\beta}=\int_0^\infty d\sigma\int\rho(dy)\,{\rm div}V_\beta(y)
	.{\rm div}V_\alpha(f^\sigma y)      $$
\indent
	In the discrete time case we define the thermodynamic flux conjugate to $E_\alpha$ as the following first order expression in $X$, analogous to (9),
$$	{\cal I}_\alpha={1\over2}\rho({\rm div}X.{\rm div}V_\alpha)
	-\delta\rho(({\rm div}V_\alpha)\circ f)      $$
$$	={1\over2}\int\rho(dy)\,{\rm div}X(y).{\rm div}V_\alpha(y)
	+\sum_{\sigma=1}^\infty\int\rho(dy)\,{\rm div}X(y)
	.{\rm div}V_\alpha(f^\sigma y)
	=\sum_\beta L_{\alpha\beta}E_\beta\eqno{(10)}      $$
where
$$	L_{\alpha\beta}={1\over2}\int\rho(dy){\rm div}V_\beta(y)
	.{\rm div}V_\alpha(y)+\sum_{\sigma=1}^\infty\int\rho(dy)
	{\rm div}V_\beta(y).{\rm div}V_\alpha(f^\sigma y)      $$
\indent
	The definitions (9) and (10) are such that the entropy production ( given by (7) or (8)) is 
$$	e=\sum_\alpha{\cal I}_\alpha E_\alpha
	=\sum_{\alpha\beta}L_{\alpha\beta}E_\alpha E_\beta      $$
\indent
	{\bf 6. An operator formulation.}
\medskip
	Let us assume that there is a (sufficiently large) Banach space ${\cal B}$ of continuous functions $\Phi:M\to{\bf R}$ such that $\rho(\Phi)=0$ and
$$	\int_{-\infty}^\infty d\sigma|\rho(\Psi.(\Phi\circ f^\sigma))|
	\le C||\Phi||_{\cal B}||\Psi||_{\cal B}\eqno{(11)}      $$
or
$$	\sum_{\sigma=-\infty}^\infty d\sigma|\rho(\Psi.(\Phi\circ f^\sigma))|
	\le C||\Phi||_{\cal B}||\Psi||_{\cal B}\eqno{(12)}      $$
if $\Phi$, $\Psi\in{\cal B}$.  In particular, (12) holds if $f$ is an Anosov diffeomorphism and ${\cal B}$ is the space of $\alpha$-H\"older continuous functions for some $\alpha>0$, with vanishing $\rho$-integral.
\medskip
	We want to take advantage of the fact the formulae for ${\cal I}_\alpha$ only involve the divergence of $X$ and $V_\alpha$ to give an operator formulation of the relation between thermodynamic forces, fluxes, and entropy production, with time dependent $X_t=\delta_tF$ or $\delta_tf\circ f^{-1}$.  Assuming that the divergence of $X_t$ is in ${\cal B}$, we define ${\cal X}_t\in{\cal B}$ and ${\cal I}_t\in{\cal B}^*$ (the dual of ${\cal B}$) as follows\footnote{*}{We give here a corrected version of (8), (8') of [46].}
$$	{\cal X}(t)=-{\rm div}X_t      $$
and
$$	{\cal I}(t)=\delta_t\rho\qquad\qquad\hbox{(continuous time)}      $$
or
$$	{\cal I}(t)=-{1\over2}({\rm div}X_t).\rho+f^*\delta_{t-1}\rho\qquad
	\qquad\hbox{(discrete time)}      $$
[In the time independent situation of Section 5 we have thus ${\cal I}_\alpha=({\cal I}(0),-{\rm div}V_\alpha)$].  The entropy production (7) or (8) may now be written
$$	e_t=({\cal I}(t),{\cal X}(t))      $$
\indent
	Let us introduce a linear operator $K(\sigma):{\cal B}\to{\cal B}^*$ such that 
$$	(K(\sigma)\Psi,\Phi)=\left\{\matrix{0\qquad\hbox{for}\qquad\sigma<0\cr
	{1\over2}\rho(\Psi\Phi)\qquad\hbox{for}\qquad\sigma=0\cr
\rho(\Psi.(\Phi\circ f^\sigma))\qquad{for}\qquad\sigma>0\cr}\right.      $$
[This is a slight modification of the definition of $\kappa_\sigma$; $K(\sigma)$ is a response function adapted to the situation close to equilibrium.  Note that the value at $\sigma=0$ is relevant only in the discrete time case].  We have now
$$	{\cal I}(t)=\int d\sigma\,K(\sigma){\cal X}(t-\sigma)      $$
or
$$	{\cal I}(t)=\sum_\sigma K(\sigma){\cal X}(t-\sigma)      $$
in the discrete and continuous time cases respectively.  Note that causality is respected in the sense that ${\cal I}(t)$ depends only on ${\cal X}(t-\sigma)$ for $\sigma\ge0$.  Let $L(\omega):{\cal B}\to{\cal B}^*$ be the real part of the Fourier transform of $K(\sigma)$:
$$	L(\omega)=\int_0^\infty d\sigma\,\cos\omega\sigma\, K(\sigma)      $$
or
$$	=\sum_{\sigma=0}^\infty\cos\omega\sigma\, K(\sigma)      $$
[The convergence of the right-hand sides is ensured by (11) or (12)].  In the time independent case (writing ${\cal I}(t)={\cal I}$, ${\cal X}(t)={\cal X}$) we have thus
$$	{\cal I}=L(0){\cal X}      $$
More generally, let ${\cal I}(\omega)$ be defined (in the continuous time case) by
$$	({\cal I}(\omega),\Phi)=\langle\int_0^\infty d\sigma\,
	\rho((-{\rm div}X_{t-\sigma}).\Phi_t\circ f^\sigma)\rangle_t      $$
where $X_t=\sqrt{2}\sin(\omega t-\theta)X$, $\Phi_t=\sqrt{2}\sin(\omega t-\theta)\Phi$, and $\langle\ldots\rangle_t$ is average over $t$; taking ${\cal X}=-{\rm div}X$ we have 
$$	{\cal I}(\omega)=L(\omega){\cal X}      $$
The discrete time case is similar.
\medskip
	{\bf 7. Onsager reciprocity and the fluctuation-dissipation theorem.}
\medskip
	We may interpret the Fourier transform $\hat K(\omega)$ of the response function $K(\sigma)$ as the {\it susceptibility}.  We have in the continuous time case
$$	(\hat K(\omega)\Psi,\Phi)=\int_0^\infty d\sigma\,e^{i\omega\sigma}
	\rho(\Psi.\Phi\circ f^\sigma)      $$
and in the discrete time case
$$	(\hat K(\omega)\Psi,\Phi)={1\over2}\rho(\Psi.\Phi)
+\sum_{\sigma=1}^\infty e^{i\omega\sigma}\rho(\Psi.\Phi\circ f^\sigma)      $$
\indent
	In this section we shall make use of {\it reversibility}.  We may thus define an operator $\epsilon:{\cal B}\to{\cal B}$ by $\epsilon\Phi=\Phi\circ i$, and we have $\epsilon^2=1$.  Also ${\cal B}={\cal B}_++{\cal B}_-$ where ${\cal B}_+=\{\Phi+\epsilon\Phi:\Phi\in{\cal B}\}$, ${\cal B}_-=\{\Phi-\epsilon\Phi:\Phi\in{\cal B}\}$.  Note that if we write $\epsilon(\Phi)=\pm1$ when $\Phi\in{\cal B}_\pm$ we have $\epsilon\Phi=\epsilon(\Phi)\Phi$.
\medskip
	{\bf Proposition.}
\medskip{\sl
	(a) $(\hat K(\omega)\Psi,\Phi)$ extends analytically to ${\rm Im}\omega>0$ (by causality).
\medskip
	(b) $(\hat K(\omega)\Psi,\Phi)=\epsilon(\Phi)\epsilon(\Psi)(\hat K(\omega)\Phi,\Psi)$ if $\Phi\in{\cal B}_\pm$, $\Psi\in{\cal B}_\pm$ (by reversibility).
\medskip
	(c) The real part $L(\omega)$ of $\hat K(\omega)$ satisfies
$$	L(\omega)={1\over2}(\hat K(\omega)+\hat K(-\omega))      $$
$$	(L(\omega)\Psi,\Phi)=\hat\rho_{\Psi\Phi}(\omega)\eqno{(13)}      $$
$$	\epsilon^*L(\omega)\epsilon=L(\omega)\eqno{(14)}      $$}
\indent
	Part (b) and equation (14) result from
$$	\rho(\Psi.\Phi\circ f^\sigma)=(i^*\rho)(\Psi.\Phi\circ f^\sigma)
	=\rho((\Psi\circ i).(\Phi\circ i\circ f^{-\sigma}))
	=\rho(\epsilon\Phi.(\epsilon\Psi)\circ f^\sigma)      $$
The other assertions are immediate.\qed
\medskip
	{\bf Remark.}
\medskip
	By (13), the quadratic form on ${\cal B}$ defined by $\Phi\mapsto(L(\omega)\Phi,\Phi)$ is $\ge0$ because $\hat\rho_{\Phi\Phi}\ge0$ as discussed in Section 4.9.
\medskip
	{\bf Reciprocity relations} (Onsager [68]).
\medskip
	Taking $\omega=0$ and writing $L(0)=L$, so that ${\cal I}=L{\cal X}$ in the time independent case, we obtain from (14) the usual reciprocity relations
$$	(L\Psi,\Phi)=\epsilon(\Phi)\epsilon(\Psi)(L\Phi,\Psi)      $$
or, with $\epsilon_\alpha=\epsilon(\Phi_\alpha)$, $\epsilon_\beta=\epsilon(\Phi_\beta)$, and the notation of Section 5
$$	L_{\alpha\beta}=\epsilon_\alpha\epsilon_\beta L_{\beta\alpha}      $$
\medskip
	{\bf Fluctuation-dissipation theorem} (Green [50], Kubo [57]).
\medskip
	The formula (13) gives 
$$	{1\over2}[(K(\omega)(-{\rm div}X),\Phi)+(K(-\omega)(-{\rm div}X),\Phi)]
	=\hat\rho_{(-{\rm div}X),\Phi}(\omega)      $$
{\it i.e.}, the real part of the susceptibility corresponding to the force $X$ is given by the Fourier transform of a time correlation function for the equilibrium state $\rho$.  This is a form of the fluctuation-dissipation theorem.  The left-hand side corresponds to dissipation (the nonequilibrium process associated with $X$) and the right-hand side to fluctuation (in the equilibrium state $\rho$).  The physical use of the above formula requires that we can identify $-{\rm div}X$ as an observable (see Section 4).  We have discussed in Section 4.10 what remains of the fluctuation-dissipation theorem far from equilibrium.
\medskip
	{\bf 8. Higher order corrections, and noisy systems.}
\medskip
	Consider the continuous time dynamical system $(f_\lambda^t)$
associated with the vector field $F+\lambda X$ on $M$.  [It is
convenient here to make the powers of $\lambda$ explicitly visible,
while in earlier sections we took $\lambda=1$.  We take $F+\lambda X$ to
be of degree 1 in $\lambda$ for simplicity].  Given an SRB state
$\rho_0$ for $(f_0^t)$ and assuming that one can perturb $\rho_0$ to an
SRB state $\rho_\lambda$ for small $|\lambda|$, we obtain (see [87]) a formal expansion of $\rho_\lambda$ in powers of $\lambda$, given by 
$$	\rho_\lambda(\Phi)=\sum_{r=0}^\infty\lambda^r\int_0^\infty d\tau_1
\ldots\int_0^\infty d\tau_r\,\rho_0(PQ^{\tau_r}P\ldots PQ^{\tau_1}\Phi)
	\eqno{(15)}      $$
where $\Phi:M\to{\bf R}$ is smooth and independent of $\lambda$, and the operators $P$, $Q^t$ are defined on smooth functions by
$$	P\Psi=X\cdot\nabla\Psi      $$
$$	Q^t\Psi=\Psi\circ f^t      $$
Formally, we can also write 
$$	\rho_\lambda(\Phi)
	=\rho_0((1-\lambda\int_0^\infty d\tau PQ^\tau)^{-1}\Phi)      $$
\indent
	The simplicity of these formulae is deceptive.  In particular, if one decomposes $X$ into stable, unstable, and neutral components to study the convergence of the multiple integrals in (15), one gets complicated expression which appear unusable in practice.  If one assumes that $\rho_0$ is an equilibrium state one recovers to leading order the results of Sections 2 and 3 for $\delta\rho$ and $e$, but the higher order contibutions can no longer be expressed solely in terms of ${\rm div}X$.
\medskip
	It is interesting at this point to discuss systems with a little bit of noise added.  We start with a probability space $(\Omega,{\bf P})$ and a continuous one-parameter group $(\theta^t)$ of transformations of $\Omega$ with respect to which ${\bf P}$ is ergodic.  A one-parameter group of transformations on $\Omega\times M$ is given such that the time $t$ map is $(\omega,x)\mapsto(\theta^t\omega,f_\omega^tx)$.  One could here again define SRB states (see the discussion of the discrete time case in Section 4.2).  We shall however take another approach, forget about hyperbolicity, and discuss the Markov case which is realized by stochastic differential equations (see for instance Arnold [2]).  Let $(f_{\omega\lambda}^t)$ be the family of diffeomorphisms obtained by integrating the stochastic differential equation (Langevin equation)
$$	{dx\over dt}=F(\theta^t\omega,x)+\lambda X(x)      $$
and define
$$	P\Psi=X\cdot\nabla\Psi      $$
$$	Q_\omega^t\Psi=\Psi\circ f_{\omega0}^t      $$
If $\mu$ is a probability measure with smooth density on $M$ we define a stationary measure $\rho_\lambda$ as a weak limit of $\int{\bf P}(d\omega)(f_{\omega\lambda}^T)^*\mu$ when $T\to\infty$.  Using the Markov property we find then formally
$$	\rho_\lambda(\Phi)=\rho_0((1-\lambda\int{\bf P}(d\omega)
	\int_0^\infty d\tau PQ_\omega^\tau)^{-1}\Phi)      $$
where 
$$	(\int{\bf P}(d\omega)Q_\omega^\tau\Phi)(x)
	=\int p^\tau(x,y)\Phi(y)dy      $$
and $p^\tau$ is the kernel of the diffusion process associated with the Langevin equation.
\medskip
	We must refer to [87] for a more detailed study of this problem, but we note the similarity of the expression obtained for the stationary measure $\rho_\lambda$ and the earlier expression obtained for the SRB measure $\rho_\lambda$ in the deterministic case.  In fact, in the zero noise limit, the coefficients of the $\lambda$ expansion for the stationary measure formally reproduce the coefficients for the SRB measure.  This is in accordance with various results on the stability of SRB states under small random perturbations for uniformly hyperbolic systems (see Kifer [55], [56]).  Since the roundoff errors in numerical simulations may be viewed as a small random noise, it is satisfactory that they respect SRB states.
\medskip
	We again refer to [87] for the study of higher order corrections for discrete time dynamical systems.

\vfill\eject
\centerline{Chapter 6. ISOKINETIC TIME EVOLUTIONS}
\centerline{AND THE DETTMANN-MORRISS PAIRING THEOREM.}
\bigskip\bigskip
	{\bf 1. Isokinetic models.}
\medskip
	Consider the evolution equation
$$	{d\over dt}\pmatrix{{\bf p}\cr{\bf q}\cr}=
\pmatrix{{\bf\xi}({\bf q})-\alpha{\bf p}\cr{\bf p}/m\cr}\eqno{(1)}$$
where ${\bf p}$, ${\bf q}\in {\bf R}^N$ and $\alpha={\bf p}\cdot{\bf\xi}/
{\bf p}\cdot{\bf p}$.  This is called an isokinetic time evolution
because it conserves the kinetic energy ${\bf p}\cdot{\bf p}/2m$,
and we may thus restrict (1) to the manifold $S\times{\bf R}^N$ where
$$	S=\{{\bf p}\in {\bf R}^N:{\bf p}\cdot{\bf p}/2m=K\}     $$
\par
	We may replace the configuration space ${\bf R}^N$ by an open
subset $M$ of ${\bf R}^N$ or of the $N$-dimensional torus ${\bf T}^N$,
with a piecewise smooth boundary $\partial M$.  The time evolution (1)
in $S\times M$ is then complemented by elastic reflection at the
boundary $\partial M$ (see Section 5).  Such a model is discussed
in [16], [17] (where ${\bf\xi}$ is taken to be the sum of an electric force
and a magnetic force).  A system of $n$ hard balls in ${\bf R}^d$ can
also be treated in this manner.  Here we take $N=nd$ and, if the balls
have different masses $m_k$, positions $x_k$ and velocities $v_k$, let
$$x_k=({m_k\over m})^{1/2}q_k\qquad,\qquad v_k=({m\over m_k})^{1/2}p_k$$
(with these variables ${\bf p}$, ${\bf q}$, the shocks between the balls
correspond to elastic reflections at the boundary $\partial M$).
\medskip
	A generalization of the isokinetic evolution (1) consists in
replacing the configuration space ${\bf R}^N$ by a manifold $M$ with a
Riemann metric (see Section 10).  In the absence of force we obtain thus
a geodesic flow on $M$.
\medskip
	Suppose that the force ${\bf\xi}$ in the isokinetic evolution
equation (1) is locally a gradient (this is the case for an electric,
but not for a magnetic force).  Then Dettmann and Morriss [22] have proved a
{\it pairing property} of Lyapunov exponents (see Theorem 7).  Such a 
pairing has been observed in the numerical study of various models (see
[21], [61]), and its rigorous justification (in [22], [23], [24], [18], [96]) is
the main reason for the interest enjoyed by isokinetic models.
\medskip
	{\bf 2. Entropy production.}
\medskip
	The divergence ${\rm div}F$ of the right-hand side of (1) is
readily computed to be $-(N-1)\alpha$ (this does not require that $\xi$
be locally a gradient).  The entropy production $e=e(\rho)$, which is equal
to minus the expectation value of ${\rm div}F$, takes thus the simple form
$$	e(\rho)=(N-1)\rho(\alpha)={N-1\over2K}w     $$
where $w=\rho({\bf p}\cdot{\bf\xi}/m)$ is the work performed on the
system (or dissipated) by unit time, and $N-1$ may be interpreted as the
number of degrees of freedom.  In equilibrium statistical mechanics the
"equipartition of energy" would give $K/(N-1)=T/2$.  (We take here
Boltzmann's constant $k=1$, as we have done for the definition of
entropy production).  Therefore
$$	e={w\over T}     $$
which fits the formula
$$	d{\bf S}={dQ\over T}     $$
where ${\bf S}$ is the entropy, but notice that here we may be far from equilibrium.
\medskip
	{\bf 3. The variations $\delta{\bf p}$, $\delta{\bf q}$.}
\medskip
	We suppose ${\bf p}\ne0$ and consider a solution $X:t\mapsto
({\bf p},{\bf q})\in S\times{\bf R}^N$ of (1), and an infinitesimally close
solution $X_\Delta:t\mapsto({\bf p}+\Delta{\bf p},{\bf q}+\Delta{\bf q})$
also in $S\times{\bf R}^N$.  Define $\delta{\bf p}=\Delta{\bf p}$ and
$\delta{\bf q}=\Delta{\bf q}-({\bf p}\cdot\Delta{\bf p}/{\bf p}\cdot{\bf p})
{\bf p}$, then ${\bf p}\cdot\delta{\bf p}={\bf p}\cdot\delta{\bf q}=0$,
and $X_\delta:t\mapsto({\bf p}+\delta{\bf p},{\bf q}+\delta{\bf q})$ is
(neglecting higher order terms) a reparametrization of $X_\Delta$.  It will
be useful to consider $X_\delta$ when we discuss Lyapunov exponents for
the flow defined by (1).
\medskip
	{\bf Lemma.}
\medskip	
	{\sl Let $({\bf e}_i)_{i=1}^{N-1}$ be an orthonormal basis
of the subspace of ${\bf R}^N$ orthogonal to ${\bf p}$, so that
$d{\bf e}_i/dt=-({\bf \xi}\cdot{\bf e}_i/{\bf p}\cdot{\bf p}){\bf
p}$.  Writing $\delta{\bf p}=\sum_{i=1}^{N-1}\delta p_i{\bf e_i}$,
$\delta{\bf q}=\sum_{i=1}^{N-1}\delta q_i{\bf e_i}$, we have}
$$	{d\over dt}\pmatrix{\delta p_i\cr\delta q_i\cr}=
\pmatrix{-\alpha\delta_{ij}&\partial_j\xi_i\cr(1/m)\delta_{ij}&0\cr}
\pmatrix{\delta p_j\cr\delta q_j\cr}\eqno{(2)}$$
\medskip
	Note that the conditions on the $d{\bf e}_i/dt$ ensure that $({\bf
e}_i)$ remains an orthonormal basis of the subspace of ${\bf R}^N$
orthogonal to ${\bf p}$ (they define parallel transport of $({\bf e}_i)$
along $X$).  We have
$$	{d\over dt}\Delta{\bf p}=(\Delta{\bf q}\cdot{\bf \partial}){\bf \xi}
	-\alpha\Delta{\bf p}-(\Delta\alpha){\bf p}     $$
$$	{d\over dt}\Delta{\bf q}={1\over m}\Delta{\bf p}     $$
We may replace $\Delta{\bf p}$, $\Delta{\bf q}$, by $\delta{\bf p}$,
$\delta{\bf q}$, in the right-hand sides ($\Delta{\bf q}$ is initially chosen
orthogonal to ${\bf p}$), then take a projection parallel to ${\bf p}$
to obtain (2).\qed
\medskip
	{\bf 4. Proposition.} (Dettmann-Morriss [22]).
\medskip	
	{\sl If $\delta x(t)$ is the vector with components
$\delta p_i$, $\delta q_i$, we may write
$$	\delta x(t)=L(t)\delta x(0)     $$
where $L(t)$ is a $(2N-2)\times(2N-2)$ matrix and $L(0)={\bf 1}$.  Also let
$$	J=\pmatrix{0&\delta_{ij}\cr-\delta_{ij}&0\cr}     $$
Then if $\xi$ is locally a gradient, we have
$$	L(t)^*JL(t)=\exp (-\int_0^t\alpha(X(s))ds)J     $$
where $L^*$ is the transpose of $L$, and $\alpha(x)={\bf p}\cdot{\bf\xi}
/{\bf p}\cdot{\bf p}={\bf p}\cdot{\bf\xi}/2K$ when $x=({\bf p},{\bf q})$.}
\medskip
	In view of (2) we have
$$	{d\over dt}L(t)=TL(t)     $$
where
$$	T=\pmatrix{-\alpha\delta_{ij}&\partial_i\xi_j\cr
	{1\over m}\delta_{ij}&0\cr}     $$
so that
$$	T^*J+JT=\pmatrix{0&-\alpha\delta_{ij}\cr
	\alpha\delta_{ij}&\partial_j\xi_i-\partial_i\xi_j\cr}     $$
If $\xi$ is locally a gradient this gives
$$	T^*J+JT=-\alpha J     $$
hence
$$	{d\over dt}[L(t)^*JL(t)]=-\alpha[L(t)^*JL(t)]     $$
and since $L(0)^*JL(0)=J$, we obtain
$$	L(t)^*JL(t)=\mu J\eqno{(3)}     $$
with $\mu=\exp(-\int_0^t\alpha(X(s))ds)$.\qed
\medskip
The formula (3) is expressed by saying that $L(t)$ is {\it conformally 
symplectic}.
\medskip
	{\bf 5. Proposition.} (Elastic reflection).
\medskip	
	{\sl Elastic reflection at an obstacle
corresponds to a linear transformation $\delta X\to\delta^RX=R\delta X$
where $R$ is symplectic ($R^*JR=J$).}
\medskip
	This result, in one form or another, is very well known.  For
completeness we give a proof.  As in the previous proposition, $\delta
X\in{\bf R}^{2N-2}$ is a vector with components $\delta p_i$, $\delta
q_i$ such that $\delta{\bf p}=\sum_i\delta p_i{\bf e}_i$,
$\delta{\bf q}=\sum_i\delta q_i{\bf e}_i$, and ${\bf p}\cdot\delta{\bf p}=
{\bf p}\cdot\delta{\bf q}=0$.  Similarly for $\delta^RX$:
$\delta^R{\bf p}=\sum_i\delta^R p_i{{\bf e}^R}_i$,
$\delta^R{\bf q}=\sum_i\delta^R q_i{{\bf e}^R}_i$.
\medskip
	Notice that $\delta^R{\bf q}$ can be computed (to first order)
as if the obstacle were planar, with normal vector ${\bf n}$, so that
$$\delta^R{\bf q}=\delta{\bf q}-2({\bf n}\cdot\delta{\bf q}){\bf n}  $$
Taking
$$	{{\bf e}^R}_i={\bf e}_i-2({\bf n}\cdot{\bf e}_i){\bf n}      $$
we have thus $\delta^R q_i=\delta q_i$ for $i=1,\ldots,N-1$.
\medskip
	To obtain $\delta^R{\bf p}$, we may assume that the surface $\Sigma$
of the obstacle is given in ${\bf R}^N$ by $x_N=\Phi(x_1,\ldots,x_{N-1})$
where $\Phi$ and its first order derivatives vanish at $(0,\ldots,0)$. 
Neglecting higher order, we see that the normal ${\bf n}+\delta{\bf n}$
to $\Sigma$ at $(x_1,\ldots,x_{N-1},\Phi(x_1,\ldots,x_{N-1}))$ is
$(-\partial_1\Phi,\ldots,-\partial_{N-1}\Phi,1)$.  We have
$$	{\bf p}^R+\delta^R{\bf p}={\bf p}+\delta{\bf p}
	-2[({\bf n}+\delta{\bf n})\cdot({\bf p}+\delta{\bf p})]
	({\bf n}+\delta{\bf n})     $$
so that
$$	\delta^R{\bf p}=\delta{\bf p}-2({\bf n}\cdot{\bf p})\delta{\bf n}
-2[({\bf n}\cdot\delta{\bf p})+({\bf p}\cdot\delta{\bf n})]{\bf n}     $$
$$	=\delta{\bf p}-2({\bf n}\cdot{\bf p})\delta{\bf n}
	+2({\bf p}\cdot\delta{\bf n}){\bf n}
-2[{\bf n}\cdot(\delta{\bf p}-2({\bf n}\cdot{\bf p})\delta{\bf n}
	+2({\bf p}\cdot\delta{\bf n}){\bf n})]{\bf n}     $$
With our choice of $({{\bf e}^R}_i)$, this means that the components
of $\delta^R{\bf p}$ with respect to $({{\bf e}^R}_i)$ are
$$	\delta^Rp_i=\delta p_i-(2({\bf n}\cdot{\bf p})\delta{\bf n}
	-2({\bf p}\cdot\delta{\bf n}){\bf n})\cdot{\bf e}_i     $$
or, writing $\delta{\bf p}^*=\sum_i\delta^Rp_i{\bf e}_i$, we have
$$	\delta{\bf p}^*=\delta{\bf p}-2(({\bf n}\cdot{\bf p})\delta{\bf n}
	-2({\bf p}\cdot\delta{\bf n}){\bf n})      $$
$$	=\delta{\bf p}-2({\bf n}\cdot{\bf p})Q\delta{\bf n}      $$
where $Q=P^{-1}$ is the inverse of the orthogonal projection $P$ of the hyperplane $\{{\bf u}:{\bf n}\cdot{\bf u}=0\}\subset{\bf R}^N$ to $\{{\bf u}:{\bf p}\cdot{\bf u}=0\}\subset{\bf R}^N$.  It is readily seen that
$$	\delta{\bf n}=-AQ\delta{\bf q}      $$
where $A$ is the matrix  if second order derivatives of $\Phi$ at 0.  Therefore
$$	\delta{\bf p}^*
	=\delta{\bf p}+2({\bf n}\cdot{\bf p})QAQ\delta{\bf q}      $$
The matrix $R$ such that $\delta^RX=R\delta X$ is thus
$$	R=\pmatrix{1&2({\bf p}\cdot{\bf n})QAQ\cr0&1}     $$
Since the matrix $A$ is symmetric, $QAQ$ is symmetric, and $R^*JR=J$, proving that $R$ is symplectic.\qed
\medskip
	{\bf 6. Lemma} (Spectrum of a conformally symplectic matrix).
\medskip
	{\sl If $L^*JL=\mu J$, then the spectrum of $L^*L$ (including
multiplicities) is invariant under $\lambda\mapsto\mu^2\lambda^{-1}$. 
The eigenspaces for the eigenvalues $\lambda$ and $\mu^2\lambda^{-1}$
are interchanged by $J$.}
\medskip
	By assumption
$$	L^*=\mu JL^{-1}J^{-1}     $$
Using also the fact that $J$ is unitary we obtain
$$	L^*L=\mu^2 JL^{-1}L^{*-1}J^{-1}=\mu^2 J(L^*L)^{-1}J^{-1}     $$
This unitary equivalence proves the lemma since  $J^{-1}=-J$ maps the
eigenspace for the eigenvalue $\lambda$ to the eigenspace for the
eigenvalue $\mu^2\lambda^{-1}$.\qed
\medskip
	{\bf 7. Theorem.} (Pairing of Lyapunov exponents).
\medskip	
	{\sl Let $M$ be an open subset of ${\bf R}^N$ or of the torus
${\bf T}^N$, with a piecewise smooth boundary $\partial M$.  Consider the
time evolution on $S\times\overline M$ defined on $S\times M$ by (1) with
$\xi$ locally a gradient, and at $\partial M$ by elastic
reflection.  Assume that $\rho$ is a probability measure on
$S\times\overline M$ such that $f^tx$ is well defined $\rho(dx)$-a.e., and
that $\rho$ is ergodic under the time evolution $(f^t)$.  Then, the $2N-1$
Lyapunov exponents associated with $(f^t)$ and $\rho$ may be labelled
$\lambda_i$ for $i=-N+1,\ldots,N-1$ in such a way that
$$	\lambda_0=0\qquad,\qquad\lambda_i+\lambda_{-i}=-\rho(\alpha)
	\qquad{\rm for}\enskip i\ne0\eqno{(4)}     $$
(with $\rho(\alpha)=\int\rho(dx)\,\alpha(x)$).}
\medskip
	There is a zero Lyapunov exponent associated with the direction
of flow, the other $2N-2$ are obtained as logarithms of the eigenvalues of
$$	\Lambda=\lim_{t\to\infty}(L(t)^*L(t))^{1/2t}     $$
for $\rho$-a.e. $x$, and $L(t)$ defined by $\delta x(t)=L(t)\delta
x(0)$.  Writing $L(t)=L$, we have by Propositions 4 and 5
$$	L^*JL=\mu J     $$
with $\mu=\exp(-\int_0^t\alpha(x(s))ds)$.  Therefore
$$	L=-\mu JL^{*-1}J     $$
$$	L^*L=\mu^2J^{-1}(L^*L)^{-1}J     $$
so that
$$	\Lambda=(\exp-\rho(\alpha))J^{-1}\Lambda^{-1}J     $$
Using Lemma 6 we see that the eigenvalues of $\Lambda$ come by pairs
$(u,v)$ with $uv=e^{-\rho(\alpha)}$; the theorem then follows
readily.\qed
\medskip
	{\bf Remark.}
\medskip
	{\sl With the above notation there are subspaces $V_{-N+1}\subset
\ldots\subset V_{-1}\subset V_1\subset\ldots\subset V_{N-1}
={\bf R}^{2N-2}$ such that
$$	\lim_{t\to\infty}{1\over t}\log\|L(t)u\|=\lambda_i\qquad
	{\rm for}\qquad u\in V_j\backslash V'_j     $$
where we have written $V'_j=V_{j-1}$ if $j\ne1$ and $V'_1=V_{-1}$.  The
spaces $V_j$ are such that for $j\ge1$, $V_{-j}$ is the skew-orthogonal $J(V'_j)^\perp$ of $V'_j$.}
\medskip
	If $U_i$ is the eigenspace of $\Lambda$ corresponding to the
eigenvalue $\exp(\lambda_i)$ we have $V_j=\oplus_{i\le j}U_i$ and, since
$U_{-j}=JU_j$,
$$	V_{-j}=\oplus_{i\ge j}U_{-i}=J\oplus_{i\ge j}U_i=J(V'_j)^\perp     $$
which establishes the asserted skew-orthogonality.\qed
\medskip
	[Basically, Theorem 7 is due to Dettmann and Morriss [22], but the
inclusion of shocks and the above remark on skew-orthogonality are due to
Wojtkowski and Liverani [96]].
\medskip
	{\bf 8. Time dependent forces.}
\medskip
	Let {\bf P} be an ergodic measure for the flow $(\theta^t)$ on
$\Omega$, and let ${\bf\xi}_\omega$ be a vector field on $M$ depending
on $\omega\in\Omega$.  Writing also $\alpha_\omega=
{\bf p}\cdot{\bf\xi}_\omega/{\bf p}\cdot{\bf p}$, we obtain a time evolution
$x(0)\mapsto f_\omega^t x(0)=x(t)=({\bf p},{\bf q})$ by solving
$$	{d\over dt}\pmatrix{{\bf p}\cr{\bf q}\cr}
=\pmatrix{{\bf\xi}_{\theta^t\omega}({\bf q})-\alpha_{\theta^t\omega}{\bf p}\cr
	{\bf p}/m\cr}     $$
on $S\times M$, with elastic collisions at $\partial M$.  Clearly
$f_\omega^{t+s}=f_{\theta^t\omega}^sf_\omega^t$, so that $(\omega,t)\mapsto
{\bf f}^t(\omega,t)=({\theta^t\omega},f_\omega^tx)$ defines a flow
$({\bf f}^t)$ on $\Omega\times S\times M$.  The formula for the entropy
production is here
$$	e({\bf\rho})=(N-1)\int{\bf\rho}(d\omega dx)\alpha_\omega(x)     $$
If ${\bf\rho}$ is an $({\bf f}^t)$-ergodic measure (with projection ${\bf
P}$ on $\Omega$) we can define Lyapunov exponents $\lambda_i$ such that
the $\exp\lambda_i$ are the eigenvalues of
$$\Lambda_{\omega,x}=\lim_{t\to\infty}((T_xf_\omega^t)^*(T_xf_\omega^t))^{1/2t}$$
for ${\bf\rho}$-almost all $(\omega,x)$.  If ${\bf\xi}_\omega$ is
locally a gradient (for ${\bf P}$-almost all $\omega$), then the
$\lambda_i$ again satisfy the pairing property (use the same arguments
as in the time independent case).  Here
$$  \lambda_i+\lambda_{-i}=-\int{\bf\rho}(d\omega dx)\alpha_\omega(x)  $$ 
\par
	{\bf 9. Generalizations: conformally symplectic structures.}
\medskip
	Let ${\cal M}$ be a $2N$-dimensional manifold.  Wojtkowski and
Liverani [96] define a conformally symplectic structure on $M$ to be a
differentiable $2$-form $\Theta$ which is non-degenerate and such that
$$	d\Theta=\gamma\wedge\Theta     $$
for some closed 1-form $\gamma$.  Locally, we have thus $\gamma=dU$ and
$d(e^{-U}\Theta)=0$.
\medskip
	Given a function $H:{\cal M}\to{\bf R}$ (Hamiltonian), we define
a vector field $F=\nabla_\Theta H$ by
$$	\Theta(\cdot,F)=dH(\cdot)\eqno{(5)}     $$
Since $e^UF$ is locally a Hamiltonian vector field with respect to the
symplectic form $e^{-U}\Theta$, the {\it conformally Hamiltonian flow}
$(f^t)$ defined by $F$ conserves the energy $H$.
\medskip
	Let ${\cal M}_K=\{x\in {\cal M}:H(x)=K\}$ and $\Theta_K$ be the
restriction of $\Theta$ to $T^{\otimes2}{\cal M}_K$.  If $u,v\in T_x{\cal
M}_K$, we write
$$({d\over dt}\Theta_K)(u,v)={d\over dt}\Theta(T_xf^tu,T_xf^tv)\big|_{t=0}$$
\medskip
	{\bf Theorem.} (Wojtkowski-Liverani [96])
\medskip
	{\sl For a conformally Hamiltonian flow restricted to the level
set ${\cal M}_K$, we have
$$	{d\over dt}\Theta_K=\gamma(F)\Theta_K     $$
where $F$ is defined by (5).}
\medskip
	On ${\cal M}_K$ we have indeed
$$	d(e^{-U}(H-K))(\cdot)=e^{-U}dH(\cdot)=e^{-U}\Theta(\cdot,F)     $$
{\it i.e.}, $F$ coincides with the Hamiltonian vector field given by the
Hamiltonian $e^{-U}(H-K)$ with respect to the symplectic form
$e^{-U}\Theta$.  Since the symplectic form is preserved by the
Hamiltonian flow we have
$$	0={d\over dt}(e^{-U}\Theta_K)
	=-e^{-U}{dU\over dt}\Theta_K+e^{-U}{d\over dt}\Theta_K     $$
where
$$	{dU\over dt}=dU(F)=\gamma(F)     $$
and the theorem follows.\qed
\medskip
	Since $\Theta(T_x{\cal M}_K,F(x))=0$, $\Theta_K$ defines a
bilinear form on $T_x{\cal M}_K/{\bf R}.F(x)$, and we may choose a
basis of this $(2N-2)$-dimensional space so that $\Theta_K$ is defined
by the matrix $J$.  Therefore the theorem implies that
$T_xf^t({\rm mod}\,{\bf R}.F)$ is given by a matrix $L(t)$ such that
$$	L(t)^*JL(t)=J.\exp\int_0^t\gamma(F(f^tx))\, dt\eqno{(6)}   $$
We shall now show how to recover Theorem 7 from (6) in a somewhat more
general situation\footnote{*}{Another application treated by Wojtkowski and
Liverani [96] is to Nos\'e-Hoover dynamics (first discussed by Dettmann
and Morriss [24]).}.
\medskip
	{\bf 10. Isokinetic dynamics on a Riemann manifold.}
\medskip
	Let $M$ be an $N$-dimensional Riemann manifold, and
$\xi=\sum_{i=1}^N{\bf\xi}_i({\bf q})dq^i$ a closed 1-form ({\it i.e.},
$(\xi_i)$ is locally a gradient).  We let ${\cal M}=T^*M$ be the
cotangent bundle of $M$; its points are of the form $({\bf p},{\bf q})$
with ${\bf q}\in M$, ${\bf p}\in T_{\bf q}^*M$.  We take as Hamiltonian
the kinetic energy
$$	H={1\over2m}\sum_{i,j}g^{ij}p_ip_j     $$
where $(g^{ij})$ is the inverse of the matrix $(g_{ij})$ defining the
metric.  Write also
$$  \kappa=\sum_ip_idq^i\qquad,\qquad\omega=d\kappa=\sum_idp_i\wedge dq^i   $$
$$	\Theta=\omega+{1\over2K}\xi\wedge\kappa     $$
Then
$$	d\Theta=-{1\over2K}\xi\wedge d\kappa=-{1\over2K}\xi\wedge\Theta     $$
{\it i.e.}, $\Theta$ is conformally symplectic.
\medskip
	With respect to the basis $(dp_i,dq^i)$, $\Theta$ and $dH$ are
given by
$$	\pmatrix{0&\delta_{ij}\cr-\delta_{ij}&[\xi_ip_j-\xi_jp_i]/2K\cr}\qquad,
\qquad\pmatrix{g^{ij}p_j/m\cr(\partial g^{jk}/\partial q^i)p_jp_k/2m\cr}     $$
so that the conformally Hamiltonian evolution equation is
$$	{d\over dt}\pmatrix{p_i\cr q^i\cr}=F
	=\pmatrix{[(g^{jk}p_jp_k)\xi_i-(g^{jk}\xi_jp_k)p_i]/2mK
	-(\partial g^{jk}/\partial q^i)p_jp_k/2m\cr g^{ij}p_j/m\cr}     $$
with the usual summation convention.  On the level set ${\cal M}_K$ this
reduces to the expected analogue of (1):
$$	{d\over dt}\pmatrix{p_i\cr q^i\cr}=\pmatrix
{\xi_i-(g^{jk}\xi_jp_k/2mK)p_i-(\partial g^{jk}/\partial q^i)p_jp_k/2m\cr
	g^{ij}p_j/m\cr}  $$
The pairing of Lyapunov exponents follows thus from (6) and Lemma 6:
$$	\lambda_0=0\qquad,\qquad\lambda_i+\lambda_{-i}=-\rho(\alpha)     $$
where $\alpha=-\gamma(F)=g^{ij}\xi_ip_j/2mK$.  If $M$ has a piecewise
smooth boundary and the dynamics is complemented by elastic shocks at
$\partial M$, the above considerations remain correct because elastic
reflection is symplectic.
\medskip
	[Pairing for isokinetic dynamics on a Riemann manifold has been
established by Choquard [18]; the above treatment follows Wojtkowski and
Liverani [96]].
\medskip
	{\bf 11. Homological formula for $\rho(\alpha)$.}
\medskip
	If ${\bf \xi}$ is locally a gradient,{\it i.e.}, if the
1-form $\xi={\bf \xi}\cdot d{\bf q}$ is closed, it defines a
cohomology class $[\xi]\in H^1(M)$.  Let $C(x,t)$ denote the curve
$\{{\bf q}(s):0\le s\le t\}$ associated with the initial condition
$x=({\bf p}(0), {\bf q}(0))$, then the limit
$$	\lim_{t\to\infty}{1\over t}\int_{C(x,t)}\omega     $$
exists for all continuous 1-forms $\omega$, and is $\rho(dx)$-a.e.
constant (by ergodicity of $\rho$).  This is a linear functional of
$\omega$, which vanishes on exact forms ({\it i.e.}, when $\omega=d\Phi$)
and defines thus a homology class $[\rho]\in H_1(M)$.  We may thus write
$$	w=2K\rho(\alpha)=\lim_{t\to\infty}{1\over t}\int_0^t
	{{\bf p}(s)\cdot{\bf\xi}({\bf q}(s))\over m}ds     $$
$$	={1\over m}\lim_{t\to\infty}{1\over t}\int_0^t
	m{d{\bf q}(s)\over ds}\cdot{\bf\xi}({\bf q}(s))ds     $$
$$	=\lim_{t\to\infty}{1\over t}\int_{C(x,t)}
	{\bf\xi}({\bf q})\cdot d{\bf q}     $$
$$	=\langle[\xi],[\rho]\rangle     $$

	{\bf 12. Thermodynamic limit.}
\medskip
	In equilibrium statistical mechanics "large systems" exhibit
{\it thermodynamic behavior}.  This means in particular that when the
total energy is fixed, the kinetic energy has only small fluctuations
around some average value, and {\it vice versa}.  It has thus been
suggested by G. Gallavotti (see [30], [40] ,[41], [42]) that for a large
system the {\it isokinetic} time evolution defined by (1) would be
equivalent to a (more physical) time evolution with fixed total energy.
\vfill\eject

\newdimen\xshift \newdimen\xwidth \newdimen\yshift
\def\ins#1#2#3{\vbox to0pt{\kern-#2pt \hbox{\kern#1pt #3}\vss}\nointerlineskip}

\def\eqfig#1#2#3#4#5{ \par\xwidth=#1pt
\xshift=\hsize \advance\xshift by-\xwidth \divide\xshift by 2
\yshift=#2pt \divide\yshift by 2 \line{\hglue\xshift \vbox to #2pt{\vfil #3
\includegraphics{#4} }\hfill\raise\yshift\hbox{#5}}}

\newwrite\figura   \def\8{\immediate\write\figura\bgroup}


\def\qed{\unskip\kern 6pt\penalty 500\raise -2pt\hbox
{\vrule\vbox to 10pt{\hrule width 4pt\vfill\hrule}\vrule}}
\centerline{Chapter 7. LOOSE ENDS AND OUTLOOK.}
\bigskip\bigskip
	{\bf 1. Frameworks for the study of equilibrium.}
\medskip
	The view of nonequilibrium statistical mechanics presented in previous chapters was based on chaotic microscopic dynamics, and centered on nonequilibrium steady states.  We have given priority to technical results (rigorous and less rigorous) over general ideology.  In particular we have restricted ourselves to classical systems, having little to say about the quantum case.  At this point however it is desirable to open up the discussion to more general possibilities.
\medskip
	We have chosen to analyze stationary nonequilibrium situations, rather than study the approach to equilibrium \`a la Boltzmann.  While maintaining this general philosophy, we shall see that the restriction to steady states is artificial and must be somewhat relaxed (Section 2).
\medskip
	To achieve both nonequilibrium and some kind of stationarity, we are led to discussing either {\it infinite systems} or {\it nonhamiltonian forces}.  We shall now briefly review these two cases, disregarding other approaches, in particular the important {\it escape rate method} of Gaspard, Nicolis and coworkers [48], [47].
\medskip
	{\bf Infinite systems.}
\medskip
	It is customary to think of a finite system $\Sigma$ interacting with infinite {\it heat reservoirs}.  The reservoirs are often described as free fields at different temperatures.  Since the interaction of the free fields with $\Sigma$ is localized, those degrees of freedom which have been influenced by $\Sigma$ conveniently move away to infinity and can be forgotten.  The case of a single reservoir is used to study approach to equilibrium (see Jak\v si\'c and Pillet [53]).  To achieve a nonequilibrium steady state, one needs at least two reservoirs:

\eqfig{150}{130}{}{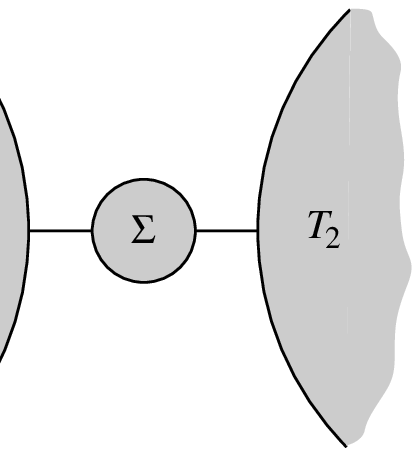}{}

The study of this situation is still in an early stage (see Rieder {\it et al.} [77], Lebowitz and Spohn [94], [63], Eckmann {\it et al.} [29], [30]).  Note that in a macroscopic description of a realistic reservoir in a steady state, the temperature $T$ would satisfy the heat equation, hence (with constant heat conductivity) $\triangle T=0$.  A limiting value $T_1$ or $T_2$ at infinity is thus possible only if the space dimension is $\ge3$.  This shows the importance of diffusive effects in nonequilibrium, and how an oversimplified model can easily give incorrect results.
\medskip
	In the infinite system situation now discussed (with Hamiltonian forces) there is no compelling reason to consider classical rather than quantum systems.  Note also that {\it dissipation} here occurs because information moves away at infinity in the reservoirs and is lost.  The study is much simplified when free fields are used as reservoirs because the degrees of freedom in the reservoirs are independent (no {\it rescattering}).  Incidentally, the absence of rescattering is also the key reason for the success of Lanford's analysis of the Boltzmann equation in the Grad limit [60].  Studying the loss of information at infinity in the presence of rescattering seems to be a formidable problem.  Since a realistic understanding of nonequilibrium in infinite systems depends on the solution of this problem, it is natural to ``cheat'' and try to understand nonequilibrium in finite systems.
\medskip
	{\bf Finite systems with nonhamiltonian forces.}
\medskip
	These are the systems studied in previous chapters.  A typical
example is the following system $\Sigma$:

\eqfig{150}{130}{}{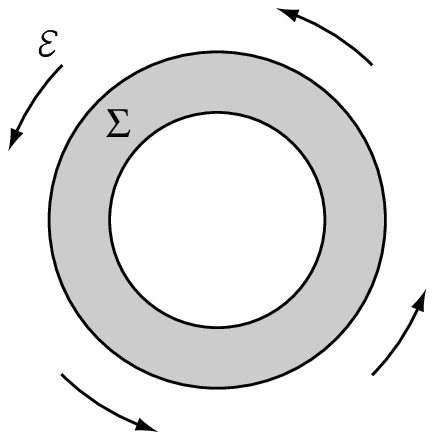}{}

$\Sigma$ consists of classical particles confined to a solid torus, and having ordinary (Hamiltonian) interactions with each other and with scatterers.  Furthermore the particles are charged and pushed around the torus by an electric field ${\cal E}$ associated locally with with a potential $U$.  In fact the electric potential is multivalued, and changed by an additive constant after one turn around the solid torus.  Finally, the system is thermostatted by fixing the total kinetic energy of the particles.  The multivalued electric potential is only mildly objectionable from the point of view of physics, and so is the isokinetic thermostat (which could be replaced by another similar thermostat).  Altogether one has a reasonably realistic physical system with the great virtue that, as we have seen, it allows a nontrivial theoretical treatment.
\medskip
	One may well argue that in statistical mechanics (nonequilibrium as well as equilibrium) the infinite system limit (thermodynamic limit) must be be taken.  We shall look into this problem in Section 3.  But note here that in equilibrium statistical mechanics it is advantageous to put the study of ensembles describing finite systems at the beginning, and leave for later the discussion of the thermodynamic limit, or the analysis of actually infinite systems.  A similar strategy seems reasonable also for nonequilibrium.
\medskip
	Unfortunately, the nice results obtained for finite classical systems with nonhamiltonian forces seem difficult to extend to the quantum situation.  Indeed, the Schr\"odinger equation requires a true potential function, not a multivalued one\footnote{*}{One can make $U$ 1-valued by going to the covering space of phase space, this approach is taken by Hoover {\it et al.} [52], Dettmann and Morriss [23].}.  One may also remark that if SRB states can forever produce entropy, it is because their own entropy is $-\infty$.  A quantum analog of this situation seems hard to achieve, because quantum entropies are positive.
\medskip
	{\bf 2. Nonsteady nonequilibrium states.}
\medskip
	We have taken as steady nonequilibrium states the limits
$$	\rho=\lim_{t\to\infty}(f^t)^*m\eqno{(1)}      $$
where $m$ is a probability measure absolutely continuous with respect to the Riemann volume element $dx$ on the phase space $M$.  But we have also asked that $\rho$ be ergodic or, better, sufficiently rapidly mixing (this is needed in applications like the linear response formula).  Of course one can always decompose an invariant state $\rho$ into ergodic components but these need not be of the form (1), or they may be of the form (1) but not mixing\footnote{**}{As an example, if the Anosov flow $(f^t)$ is not mixing, {\it i.e.}, is a suspension of an Anosov diffeomorphism $\tilde f$, one can write 
$$	\rho=\int_0^Td\tau\,(f^\tau)^*\tilde\rho      $$
where $\tilde\rho$ is an $\tilde f$-ergodic measure on a manifold $\tilde M$ with $\dim\tilde M=\dim M-1$.  Here $(f^\tau)^*\tilde\rho$ depends periodically on $\tau$.}.
\medskip
	It is not clear how to address this problem, which appears related to the question of macroscopic oscillations of nonequilibrium systems.  Indeed if a macroscopic system is maintained out of equilibrium by time independent forces it is not uncommon that it exhibits oscillations, which may be periodic or nonperiodic (chemical oscillations, hydrodynamic turbulence).  A time independent state $\rho$ describing such an oscillating system is expected to have a natural decomposition
$$	\rho=\int\mu(d\lambda)\tilde\rho_\lambda      $$
where $(f^t)^*\tilde\rho_\lambda=\tilde\rho_{\theta^t\lambda}$ and $\mu$ is ergodic under $(\theta^t)$.  It is however not clear how to express such a decomposition at the microscopic level.
\medskip
	{\bf 3. Thermodynamic limit.}
\medskip
	Statistical mechanics deals mostly with macroscopic systems, and some important phenomena like phase transitions make sense only in the large system limit.  At a technical level, an important feature of the large system limit (thermodynamic limit) in equilibrium statistical mechanics is the {\it equivalence of ensembles}.  Roughly speaking, this corresponds to the fact that for large systems, extensive variables {\it usually} fluctuate very little.  One example, which will be of interest to us, is that if one fixes the (total) energy, then the (total) kinetic energy fluctuates very little (with respect to the microcanonical ensemble, {\it i.e.}, the normalized Liouville measure on the energy shell).  This is really a fact about geometry in a large number of dimensions\footnote{**}{A typical fact of this sort is that in many dimensions, most of the mass (=Lebesgue volume) of a solid sphere is concentrated near its surface.} which is reasonably well understood in the situation of equilibrium statistical mechanics, thanks to the convexity properties of the entropy (see Ruelle [78], Lanford [59]).  The notion that extensive variables often do not fluctuate much in nonequilibrium statistical mechanics (with the microcanonical ensemble replaced by the SRB measure) probably has some validity, but a firm theoretical basis is lacking.  For a physical discussion see the papers of Evans and Sarman [33]\footnote{*}{This paper gives a perturbative treatment of the equivalence of the isokinetic and isoenergetic constraints, but ignores the singular character of the measures to be used.}, Cohen and Rondoni [19].
\medskip
	There are also some danger signs coming from physics.  Suppose for instance that we try to produce a nonequilibrium steady state corresponding to homogeneous shear in a fluid.  Macroscopic fluid theory tells us that the homogeneous shear flow that we try to produce is unstable and will become turbulent with, presumably, fluctuations on all scales. 
\medskip
	{\bf 4. Dimensionality loss and the definition of entropy.}
\medskip
	Let $\lambda_1\ge\lambda_2\ge\ldots\ge\lambda_D$ be the Lyapunov exponents associated with the SRB measure $\rho$.  If $d$ is the largest integer such that 
$$	\sum_{j=1}^d\lambda_j\ge0      $$
the {\it Lyapunov dimension} is defined to be 
$$	D_L=d+{\sum_{j=1}^d\lambda_j\over|\lambda_{d+1}|}      $$
by Kaplan and Yorke [54], [35].  According to these authors, $D_L$ is in many cases equal to the Hausdorff dimension of $\rho$.
\medskip
	For a macroscopic system outside of equilibrium it is expected that the Lyapunov exponents have an asymptotic density $\phi(\cdot)$ when $D$ (the total dimension of phase space) tends to infinity.  The number of Lyapunov exponents in the interval $(\lambda_1,\lambda_2)$ is thus $D\int_{\lambda_1}^{\lambda_2}\phi(\lambda)d\lambda$.  It is usually assumed that the support of $\phi(\cdot)$ is an interval $[\lambda_{\rm min},\lambda_{\rm max}]$ where $\lambda_{\rm min}$, $\lambda_{\rm max}$ remain finite for $D\to\infty$ (in fact $\lambda_{\rm min}$, $\lambda_{\rm max}$ may diverge exponentially, see Searles {\it et al.} [89]).  The entropy production is 
$$	e=-D\int_{\lambda_{\rm min}}^{\lambda_{\rm max}}\phi(\lambda)
	\lambda d\lambda\eqno({2})      $$
For a Hamiltonian system we have $\phi(\lambda)=\phi(-\lambda)$, while if pairing holds we have $\phi(-\lambda-\alpha)=\phi(\lambda)$ where $D\alpha/2=e$.
\medskip
	To obtain the Lyapunov dimension $D_L$ of $\rho$, we determine $\lambda_0$ such that 
$$	\int_{\lambda_0}^{\lambda_{\rm max}}\phi(\lambda)\lambda\, 
	d\lambda=0\eqno{(3)}       $$
and write 
$$    D_L=D\int_{\lambda_0}^{\lambda_{\rm max}}\phi(\lambda)\,d\lambda    $$
The dimension loss is thus
$$	\Delta=D-D_L=D\delta      $$
with 
$$	\delta
=\int_{\lambda_{\rm min}}^{\lambda_0}\phi(\lambda)\,d\lambda\eqno{(4)}      $$ 
Note that $\Delta$ is an {\it extensive variable} ({\it i.e.}, proportional to the number of degrees of freedom of the system).  From (2) and (3) we obtain 
$$	-{e\over D}=\int_{\lambda_{\rm min}}^{\lambda_0}\phi(\lambda)\lambda\, 
	d\lambda      $$
Suppose that ${\lambda_{\rm min}}$ and ${\lambda_{\rm max}}$ remain finite when $D\to\infty$.  Close to equilibrium we have $\lambda_0\approx\lambda_{\rm min}\approx-\lambda_{\rm max}$ hence (4) gives 
$$	-{e\over D}=-\lambda_{\rm max}\delta      $$
and therefore 
$$	\Delta=D\delta\approx{e\over\lambda_{\rm max}}      $$
(see Posch and Hoover [73]).
\medskip
	Corresponding to a nonequilibrium steady state $\rho$ with $\Delta>0$ we have a zero $D$-volume of occupied phase space, and therefore an entropy $-\infty$.  We may however try to define a ``Lyapunov entropy'' corresponding to $D_L$-volume in phase space.  Suppose thus that $V(\epsilon,d)$ is the occupied volume in phase space\footnote{*}{In the case of $n$ identical particles we include as usual a factor $1/n!$} in dimension $d$ associated with the length scale $\epsilon$.  We refrain from trying to give a precise definition of $V(\epsilon,d)$, but we assume a scaling form 
$$	V(\epsilon,d)\approx C\epsilon^{d-D_L}      $$
for small $\epsilon$ (such a form is familiar, for instance from renormalization group theory).  Therefore the specific entropy
$$	{1\over D}\log V(\epsilon,d)\approx{1\over D}\log C
	+{d-D_L\over D}\log\epsilon      $$
depends weakly on $\epsilon$ if $d$ is close to $D_L$, and the ``specific'' Lyapunov entropy $D^{-1}\log C$ is reasonably well defined if one takes $\epsilon^2$ of the order of Planck's constant.
\medskip
	One way to express the above formulae is that in our nonequilibrium situation a fraction $\delta$ of the degrees of freedom is frozen, and that the Lyapunov entropy $S_L=\log C$ is the entropy corresponding to the fraction $1-\delta$ that is not frozen.  But can one give a physical meaning to $S_L$? 
\medskip
	We have up to now considered classical systems.  Let us now assume for the sake of the discussion that our description is a classical approximation applied to a quantum system.  According to this bold assumption, the ``Lyapunov entropy'' $S_L$ is now the logarithm of a number of available degrees of freedom\footnote{*}{The quantum description fixes a unit of volume in phase space, based on Planck's constant, and eliminates the usual arbitrary additive constant in the definition of the entropy.}, and is thus in principle a physically meaningful quantity.  (Note that in the quantum accounting, the frozen degrees of freedom make an additive contribution equal to 0, which can be ignored, while in the classical situation, the additive constant is $-\infty$, which cannot be forgotten).
\vfill\eject
\par
{\bf References.}
\medskip

[1] L. Andrey.  ``The rate of entropy change in non-Hamiltonian systems.''  Phys. Letters {\bf 11A},45-46(1985).

[2] L. Arnold.  {\it Random dynamical systems.}  Springer, Berlin, to appear. 

[3] J. Bahnm\"uller and P.-D. Liu.  ``Characterization of measures satisfying Pesin's entropy formula for random dynamical systems.''  J. Dynam. Diff. Equa. {\bf 10},425-448(1998).

[4] V. Baladi.  ``Periodic orbits and dynamical spectra (survey).''  Ergod. Th. and Dynam. Syst. {\bf 18},255-292(1998).

[5] L. Barreira, Ya. Pesin and J. Schmeling.  ``Dimension of hyperbolic measures -- a proof of the Eckmann-Ruelle conjecture.''  Ann. of Math., to appear.

[6] P.Billingsley.  {\it Ergodic theory and information.}  John Wiley, New York, 1965.

[7] F. Bonetto and G. Gallavotti.  ``Reversibility, coarse graining and the chaoticity principle.''  Commun . Math. Phys. {\bf 189},263-276(1997).

[8] F. Bonetto, G. Gallavotti and P.L. Garrido.  ``Chaotic principle: an experimental test.''  Physica D {\bf 105},226-252(1997).

[9] R. Bowen.  ``Markov partitions for Axiom A diffeomorphisms.''  Amer. J. Math. {\bf 92},725-747(1970).

[10] R. Bowen.  ``Periodic points and measures for Axiom A diffeomorphisms.''  Trans. Amer. Math. Soc. {\bf 154},377-397(1971).

[11] R. Bowen.  ``Periodic orbits for hyperbolic flows.''  Amer. J. Math. {\bf 94},1-30(1972).

[12] R. Bowen.  ``Symbolic dynamics for hyperbolic flows.''  Amer. J. Math. {\bf 95},429-460(1973).

[13] R. Bowen.  {\it Equilibrium states and the ergodic theory of Anosov diffeomorphisms.} Lecture Notes in Math. {\bf 470}, Springer, Berlin,
1975.

[14] R.Bowen and D.Ruelle.  ``The ergodic theory of Axiom A flows.''  Invent. Math. {\bf 29},181-202(1975).

[15] N. Chernov.  ``Decay of correlations and dispersing billiards.''  J. Statist. Phys., to appear.

[16] N.I. Chernov, G.L. Eyink, J.L. Lebowitz, and Ya.G. Sinai.  ``Derivation of Ohms law in a deterministic mechanical model.''  Phys. Rev. Letters {\bf 70}, 2209-2212(1993).

[17] N.I. Chernov, G.L. Eyink, J.L. Lebowitz, and Ya.G. Sinai.  ``Steady-state electrical conduction in the periodic Lorentz gas.''  Commun. Math. Phys. {\bf 154},569-601(1993).

[18] Ph. Choquard.  ``Lagrangian formulation of Nos\'e-Hoover and of isokinetic dynamics.''  ESI report (1996).

[19] E.G.D. Cohen and L. Rondoni.  ``Note on phase space contraction and entropy production in thermostatted Hamiltonian systems.''  Chaos {\bf 8},357-365(1998).

[20] S.R. de Groot and P. Mazur.  {\it Nonequilibrium thermodynamics.}  Dover, New York, 1984.

[21] C.P. Dellago, H.A. Posch, and W.G. Hoover.  ``Lyapunov instability in a system of hard disks in equilibrium and nonequilibrium steady states.''   Phys. Rev. E {\bf 53},1483-1501(1996).

[22] C.P. Dettmann and G.P. Morriss.  ``Proof of Lyapunov exponent pairing for systems at constant kinetic energy.''  Phys. Rev. E {\bf 53},R5541-5544(1996).

[23] C.P. Dettmann and G.P. Morriss.  ``Hamiltonian formulation of the Gaussian isokinetic thermostat.''  Phys. Rev. E {\bf 54},2495-2500(1996).

[24] C.P. Dettmann and G.P. Morriss.  ``Hamiltonian reformulation and pairing of Lyapunov exponents for Nos\'e-Hoover dynamics.''  Phys. Rev. E {\bf 55},3693-3696(1997).

[25] D. Dolgopyat.  ``On decay of correlations in Anosov flows.''  Ann. of Math. {\bf 147},357-390(1998).

[26] D. Dolgopyat.  ``Prevalence of rapid mixing for hyperbolic flows.''  Ergod. Th. and Dynam. Syst. {\bf 18},1097-1114(1998)

[27] D. Dolgopyat and M. Pollicott.  ``Addendum to `Periodic orbits and dynamical spectra'.''  Ergod. Th. and Dynam. Syst. {\bf 18},293-301(1998). 

[28] J.R. Dorfman.  {\it An introduction to chaos in non-equilibrium statistical mechanics.}  Springer, Berlin, to appear.

[29] J.-P. Eckmann, C.-A. Pillet and L. Rey-Bellet.  ``Non-equilibrium statistical mechanics of anharmonic chains coupled to two heat baths at different temperatures.''  Commun. Math. Phys., to appear.

[30] J.-P. Eckmann, C.-A. Pillet and L. Rey-Bellet.  ``Entropy production in non-linear, thermally driven Hamiltonian systems.''  Preprint.

[31] D.J. Evans, E.G.D. Cohen, and G.P.Morriss.  ``Probability of second law violations in shearing steady states.''  Phys. Rev. Letters {\bf 71},2401-2404(1993).

[32] D.J. Evans and G.P. Morriss.  {\it Statistical mechanics of nonequilibrium fluids.}  Academic Press, New York, 1990.

[33] D.J. Evans and S. Sarman.  ``Equivalence of thermostatted nonlinear responses.''  Phys. Rev. {\bf E 48},65-70(1993).

[34] J. Franks and R.F. Williams.  ``Anomalous Anosov flows.'' pp. 158-174 in Lecture Notes in Math. {\bf 819}, Springer, Berlin, 1980.

[35] P. Frederickson, J.L. Kaplan, E.D. Yorke and J.A. Yorke.  ``The Lyapunov dimension of strange attractors.''  J. Diff. Equ. {\bf 49},185-207(1983).

[36] G. Gallavotti.  ``Reversible Anosov diffeomorphisms and large deviations.''  Math. Phys. Electronic J. {\bf 1},1-12(1995).

[37] G. Gallavotti.  ``Extension of Onsager's reciprocity to large fields
and the chaotic hypothesis.''  Phys Rev. Letters {\bf 77},4334-4337(1996).

[38] G. Gallavotti.  ``Chaotic hypothesis: Onsager reciprocity and fluctuation-dissipa\-tion theorem.''  J. Statist. Phys. {\bf 84},899-926(1996).

[39] G. Gallavotti.  ``Equivalence of dynamical ensembles and Navier-Stokes equations.''  Phys. Letters {\bf 223},91-95(1996).

[40] G. Gallavotti.  ``Dynamical ensembles equivalence in fluid mechanics.''  Physica D {\bf 105},163-184(1997).

[41] G. Gallavotti.  ``Chaotic dynamics, fluctuations, non-equilibrium ensembles.''  Chaos {\bf 8},384-392(1998).

[42] G. Gallavotti.  ``New methods in nonequilibrium gases and fluids.''  (Proceedings of the conference {\sl Let's face chaos through nonlinear dynamics}, U. of Maribor, 1996, ed. M. Robnik,)  Open Systems and Information Dynamics, to appear.

[43] G. Gallavotti.  ``Fluctuation patterns and conditional reversibility in nonequilibrium systems.''  Ann. Inst. H. Poincar\'e, to appear (chao-dyn@xyz.lanl.gov \#9703007).

[44] G. Gallavotti and E.G.D. Cohen.  ``Dynamical ensembles in nonequilibrium statistical mechanics.''  Phys. Rev. Letters {\bf 74},2694-2697(1995).

[45] G. Gallavotti and E.G.D. Cohen.  ``Dynamical ensembles in stationary states.'' J. Statist. Phys. {\bf 80},931-970(1995).

[46] G. Gallavotti and D. Ruelle.  ``SRB states and nonequilibrium statistical mechanics close to equilibrium.''  Commun. Math. Phys. {\bf 190},279-281(1997).

[47] P. Gaspard and J.R. Dorfman.  Phys. Rev. {\bf E 52},3525-3552(1995).

[48] P. Gaspard and G. Nicolis.  ``Transport properties, Lyapunov exponents, and entropy per unit time.''  Phys. Rev. Lett. {\bf 65},1693-1696(1990).

[49] G. Gentile.  ``Large deviation rule for Anosov flows.''  Forum Math. {\bf 10},89-118(1998).

[50] M.S. Green.  ``Brownian motion in a gas of noninteracting molecules.''  J. chem. Phys. {\bf 19},1036-1046(1951).

[51] W.G. Hoover.  {\it Molecular dynamics}.  Lecture Notes in Physics {\bf 258}.  Springer, Heidelberg, 1986.

[52] W.G. Hoover, B. Moran, C. Hoover and W.J. Evans.  ``Irreversibility in the Galton board via conservative classical and quantum Hamiltonian and Gaussian dynamics.''  Phys. Letters {\bf 133},114-120(1988).

[53] V. Jak\v si\'c and C.-A. Pillet.  ``Ergodic properties of classical dissipative systems I.''  Acta mathematica, to appear.

[54] J.L. Kaplan and J.A.Yorke.  ``Preturbulence: a regime observed in a fluid flow of Lorenz.''  Commun. Math. Phys. {\bf 67},93-108(1979)

[55] Yu. Kifer.  {\it Ergodic theory of random transformations.}  Birkh\"auser, Boston, 1986. 

[56] Yu. Kifer.  {\it Random perturbations of dynamical systems.}  Birkh\"auser, Boston, 1988.

[57] R. Kubo.  ``Statistical-mechanical theory of irreversible processes. I.''  J. Phys. Soc. (Japan) {\bf 12},570-586(1957).

[58] J. Kurchan.  ``Fluctuation theorem for stochastic dynamics.''  J. Phys. A: Math. Gen. {\bf 31},3719-3729(1998).

[59] O.E. Lanford.  ``Entropy and equilibrium states in classical statistical mechanics.'' pp. 1-113 {\it in Lecture Notes in Physics} {\bf 20}, Springer, Berlin, 1973.

[60] O.E. Lanford.  ``Time evolution of large classical systems.''  pp. 1-111 in Lecture Notes in Physics. {\bf 38}, Spinger, Berlin, 1975.

[61] A. Latz, H. van Beijeren, and J.R. Dorfman.  ``Lyapunov spectrum and the conjugate pairing rule for a thermostatted random gas: kinetic theory.''  Phys. Rev. Letters {\bf 78},207-210(1997).

[62] J.L. Lebowitz.  ``Boltzmann's entropy and time's arrow.''  Physics Today {\bf 46}, No 9,32-38(1993).

[63] J.L. Lebowitz and H. Spohn.  ``Transport properties of the Lorentz gas: Fourier's law.''  J. Statist. Phys. {\bf 19},633-654(1978).

[64] J.L. Lebowitz and H. Spohn.  ``A Gallavotti-Cohen type fluctuation theorem for stochastic dynamics.''  Preprint.

[65] F. Ledrappier.  ``Propri\'et\'es ergodiques des mesures de Sinai.''  Publ. math. IHES {\bf 59},163-188(1984).

[66] F. Ledrappier and J.-M. Strelcyn.  "A proof of the estimation from below in Pesin's entropy formula."  Ergod. Th. and Dynam. Syst. {\bf 2},203-219(1982).

[67] F.Ledrappier and L.S.Young.  "The metric entropy of diffeomorphisms: I. Characterization of measures satisfying Pesin's formula, II. Relations between entropy, exponents and dimension."  Ann. of Math. {\bf 122},509-539,540-574(1985).

[68] L. Onsager.  ``Reciprocal relations in irreversible processes. II.''  Phys. Rev. {\bf 38},2265-2279(1931).

[69] V.I. Oseledec.  ``A multiplicative ergodic theorem.  Lyapunov characteristic numbers for dynamical systems.''  Tr. Mosk. Mat. Ob\v s\v c. {\bf 19},179-210(1968).  English translation, Trans. Moscow Math. Soc. {\bf 19},197-221(1968).

[70] W. Parry and M. Pollicott.  {\it Zeta functions and the periodic orbit structure of hyperbolic dynamics.}  Ast\'erisque {\bf 187-188}, Soc. Math. de France, Paris, 1990.

[71] Ya.B.Pesin.  ``Invariant manifold families which correspond to non-vanishing characteristic exponents.''  Izv. Akad. Nauk SSSR Ser. Mat. {\bf 40},No 6,133
2-1379(1976).  English translation: Math. USSR Izv. {\bf 10},No 6,1261-1305(1976).

[72] Ya.B.Pesin.  ``Lyapunov characteristic exponents and smooth ergodic theory.''  Uspehi Mat. Nauk {\bf 32},No 4,55-112(1977).  English translation:  Russian Math. Surveys. {\bf 32},No 4,55-114(1977).

[73] H.A. Posch and W.G. Hoover.  ``Lyapunov instability in dense Lennard-Jones fluids.''  Phys. Rev. {\bf A 38},473-482(1988).

[74] I. Prigogine.  {\it Introduction to thermodynamics of irreversible processes.}  John Wiley, New York, 1962.

[75] C.C. Pugh and M. Shub.  ``Ergodic attractors.''  Trans. Amer. Math. Soc. {\bf 312},1-54(1989).

[76] M. Ratner.  ``Markov partitions for Anosov flows on 3-dimensional manifolds.''  Mat. Zam. {\bf 6},693-704(1969).

[77] Z. Rieder, J.L. Lebowitz, and E. Lieb.  ``Properties of a harmonic
crystal in a stationary nonequilibrium state.''  J. Math. Phys. {\bf 8},1073-1078(1967).

[78] D. Ruelle.  ``Correlation functionals.''  J. Math. Phys. {\bf 6},201-220(1965).

[79] D.Ruelle.  ``A measure associated with Axiom A attractors.''  Am. J. Math. {\bf 98},619-654(1976).

[80] D.Ruelle.  ``An inequality for the entropy of differentiable maps.''  Bol. Soc. Bras. Mat. {\bf 9},83-87(1978).

[81] D. Ruelle.  {\it Thermodynamic formalism.}  Addison-Wesley, Reading (Mass.),1978.

[82] D.Ruelle.  ``Ergodic theory of differentiable dynamical systems.''  Publ. Math. IHES {\bf 50},27-58(1979).

[83] D. Ruelle.  ``One-dimensional Gibbs states and Axiom A diffeomorphisms.''  J. Differential Geometry. {\bf 25},117-137(1987).

[84] D. Ruelle.  ``Resonances for Axiom A flows.''  J. Differential Geometry. {\bf 25},99-116(1987).

[85] D. Ruelle.  ``Positivity of entropy production in nonequilibrium statistical mechanics.''  J. Satist. Phys. {\bf 85},1-23(1996).

[86] D. Ruelle.  ``Differentiation of SRB states.''  Commun. Math. Phys. {\bf 187},227-241(1997).

[87] D. Ruelle.  ``Nonequillibrium statistical mechanics near equilibrium: computing higher order terms.''  Nonlinearity {\bf 11},5-18(1998).

[88] D. Ruelle.  ``General linear response formula in statistical mechanics, and the fluctuation-dissipation theorem far from equilibrium.''  Phys. Letters {\bf A 245},220-224\ (1998).

[89] D.J. Searles, D.J. Evans and D.J. Isbister.  ``The number dependence of the maximum Lyapunov exponent.''  Physica {\bf A 240},96-104(1977).

[90] Ya.G. Sinai.  ``Markov partitions and C-diffeomorphisms.''  Funkts. Analiz i Ego Pril. {\bf 2}, No {\bf 1},64-89(1968).  English translation, Functional Anal. Appl. {\bf 2},61-82(1968).

[91] Ya.G. Sinai.  ``Constuction of Markov partitions.''  Funkts. Analiz i Ego Pril. {\bf 2}, No {\bf 3},70-80(1968).  English translation, Functional Anal. Appl. {\bf 2},245-253(1968).

[92] Ya.G. Sinai.  ``Gibbsian measures in ergodic theory.''  Uspehi Mat. Nauk {\bf 27}, No {\bf 4},21-64(1972).  English translation, Russian Math. Surveys {\bf 27}, No {\bf 4},21-69(1972).

[93] S. Smale.  ``Differentiable dynamical systems.''  Bull. AMS {\bf 73},747-817(1967).  

[94] H. Spohn and J.L. Lebowitz.  ``Stationary non-equilibrium states of infinite harmonic systems.''  Commun. Math. Phys. {\bf 54},97-120(1977).

[95] M. Viana.  ``Multidimensional nonhyperbolic attractors.''  Publ. Math. IHES {\bf 85},63-96(1997).

[96] M.P. Wojtkowski and C. Liverani.  ``Conformally symplectic dynamics and symmetry of the Lyapunov spectrum.''  Commun. Math. Phys. {\bf 194},47-60(1998).

[97] L.-S. Young.  ``Statistical properties of dynamical systems with some hyperbolicity.''  Ann. of Math. {\bf 147},585-650(1998).

[98] L.-S. Young.  ``Recurrence times and rates of mixing.''  Preprint.

[99] L.-S. Young.  ``Ergodic theory of chaotic dynamical systems.''  Lecture at the International Congress of Mathematical Physics 1997.

\end